# The Capacity Region of the Wireless Ergodic Fading Interference Channel with Partial CSIT to Within One Bit

Reza K. Farsani[1]
Email: reza_khosravi@alum.sharif.ir

***Abstract:*** **Fundamental capacity limits are studied for the two-user wireless ergodic fading IC with partial Channel State Information at the Transmitters (CSIT) where each transmitter is equipped with an arbitrary deterministic function of the channel state (this model yields a full control over how much state information is available). One of the main challenges in the analysis of fading networks, specifically multi-receiver networks including fading ICs, is to obtain efficient capacity outer bounds. In this paper, a novel capacity outer bound is established for the two-user ergodic fading IC. For this purpose, by a subtle combination of broadcast channel techniques (i.e., manipulating mutual information functions composed of vector random variables by Csiszar-Korner identity) and genie-aided techniques, first a single-letter outer bound characterized by mutual information functions including some auxiliary random variables is derived. Then, by novel arguments the derived bound is optimized over its auxiliaries only using the entropy power inequality. Besides being well-described, our outer bound is efficient from several aspects. Specifically, it is optimal for the fading IC with uniformly strong interference. Also, it is sum-rate optimal for the channel with uniformly mixed interference. More importantly, it is proved that when each transmitter has access to any amount of CSIT that includes the interference to noise ratio of its non-corresponding receiver, the outer bound differs by no more than one bit from the achievable rate region given by Han-Kobayashi scheme. This result is viewed as a natural generalization of the ETW "to within one bit" capacity result for the static channel to the wireless ergodic fading case.**

## I. INTRODUCTION

Two fundamental characteristics of any multi-user wireless communication network are the interference effect due to concurrent communications from uncoordinated users and the time-varying nature due to the mobility of users. The Gaussian fading Interference Channel (IC) is a basic information theoretic model that simultaneously includes both these characteristics. It is indeed a useful model for many practical applications including ad-hoc wireless networks [1]. In recent years, many papers have been devoted to study capacity bounds for interference channels (see [2-6] and literature therein for previous and recent results). Nonetheless, even for the simple two-user IC the capacity region is still unknown in general. In 2008, a significant progress was made by Etkin, Tse and Wang (ETW) [7] that characterize the capacity region of the two-user Gaussian static (time-invariant) IC to within one bit. Following up this paper, approximate capacity results were also derived for the Gaussian IC with various types of user cooperation [8-10]. However, despite its great importance from practical viewpoints, there exist very few results for wireless fading ICs. The two-user ergodic fading IC with perfect state information at all terminals is considered in [1] and [11, 12]. In [1], capacity inner and outer bounds are analyzed and the corresponding power allocation polices are discussed. In [11], the capacity region for the channel with uniformly strong interference and also the sum-rate capacity for the channel with uniformly mixed interference is established. It is also shown that separable coding across fading states is not optimal for the two-user fading IC in general, a result which had been previously obtained in [13] for fading ICs with more than two users. The ergodic Z-interference channel (one receiver observes interference-free signal) with no CSIT is studied in [14]. Outage analysis of the two-user Gaussian fading IC is given in [15]. Fading ICs with more than two users have been also studied in literature, however, mostly relying on interference alignment techniques [16-17] and with the purpose of understanding the behavior of capacity in high signal to noise regime.

---

[1]Reza K. Farsani was with the department of electrical engineering, Sharif University of Technology. He is by now with the school of cognitive sciences, Institute for Research in Fundamental Sciences (IPM), Tehran, Iran.

In this paper, we study capacity limits for the two-user wireless ergodic fading IC with partial Channel State Information at the Transmitter (CSIT). Here, the partial side information at each transmitter is given by a deterministic function (potentially discrete-valued) of channel state. This model for CSIT, which was previously considered in [18-19] for the Gaussian fading MAC, yields a full control over how much state information is available at a given transmitter. Specifically, it unifies the cases where the transmitters have perfect or no knowledge of channel state information. Under such a general model for CSIT, we develop capacity bounds for the channel. A capacity inner bound is directly given based on the Han-Kobayashi message splitting scheme [20]. One of the main challenges in the analysis of fading networks, specifically multi-receiver networks including fading ICs, is to establish efficient capacity outer bounds. For the Gaussian static (non-fading) IC, several outer bounds have been developed in recent years [7, 21-26]. However, the fact is that the approach of the latter papers is not useful for the Gaussian time-varying fading IC for two reasons. Firstly, each of the outer bounds of [7, 21-26] is established by making the assumption that the static channel is in the weak or the mixed interference regime (for the strong interference regime the capacity region is known) while for the fading case, channel gains vary by time and therefore during the time the channel may alter among weak, mixed, or strong interference regimes. Secondly, in the derivation of the outer bounds of [7, 21-26], mutual information functions including vector random variables are directly evaluated using tools such as worst case additive noise lemma [27] or an extremal inequality given in [28]. A direct application of such techniques for the fading channel, in which the channel coefficients vary by time and the inputs depend on CSITs, is either impossible or involved in too complex computations. In fact, even for the Gaussian static channel with fixed gains, the outer bounds derived in [21-26] have very complex characterizations.

In this paper, we present a novel approach to establish an efficient outer bound for the Gaussian fading IC. In our approach, neither the worst case additive noise lemma [27] nor the extremal inequality [28] are used. Instead, by a subtle combination of broadcast channel techniques (i.e., manipulating mutual information functions composed of vector random variables by Csiszar-Korner identity) and genie-aided techniques, first we derive a single-letter outer bound characterized by mutual information functions including some auxiliary random variables. Then, by novel arguments the derived bound is optimized over its auxiliaries only using the entropy power inequality. In fact, for this latter step, we follow the approach developed in our concurrent paper [29] to evaluate the so-called UV-outer bound for the two-user fading broadcast channel. Besides being well-described, we demonstrate that our outer bound (which holds for the fading IC with any arbitrary amount of CSIT) is efficient from several aspects. Specifically, it is optimal for the fading channel with uniformly strong interference. Also, it is sum-rate optimal for the channel with uniformly mixed interference. More importantly, it is proved that when each transmitter has access to any amount of CSIT that includes the interference to noise ratio of its non-corresponding receiver, the outer bound differs by no more than one bit from the achievable rate region given by Han-Kobayashi scheme. This result is viewed as a natural generalization of the ETW "to within one bit" capacity result for the static channel to the wireless ergodic fading case.

In the following section, we present preliminaries and channel model definitions. Our main results are given in Section III.

# II. Preliminaries and Definitions

In this paper, we use the following notations: Random Variables (RV) are denoted by upper case letters (e.g. $X$) and lower case letters are used to show their realization (e.g. $x$). The Probability Distribution Function (PDF) of $X$ is denoted by $P_X(x)$ and the conditional PDF of $X$ given $Y$ is denoted by $P_{X|Y}(x|y)$. The probability of the event $A$ is expressed by $Pr(A)$. The notation $\mathbb{E}[.]$ indicates the expectation operator. The set of real numbers, nonnegative real numbers, and complex numbers are denoted by $\mathbb{R}, \mathbb{R}_+$, and $\mathbb{C}$, respectively. Given a statement $F$, the indicator function $\mathbb{1}(F)$ is equal to one if $F$ is true and zero otherwise. Finally, the function $\psi(x)$ is defined as: $\psi(x) \equiv \log(1+x)$, for $x \in \mathbb{R}_+$.

***The Two-User Gaussian Fading Interference Channel (GFIC):***

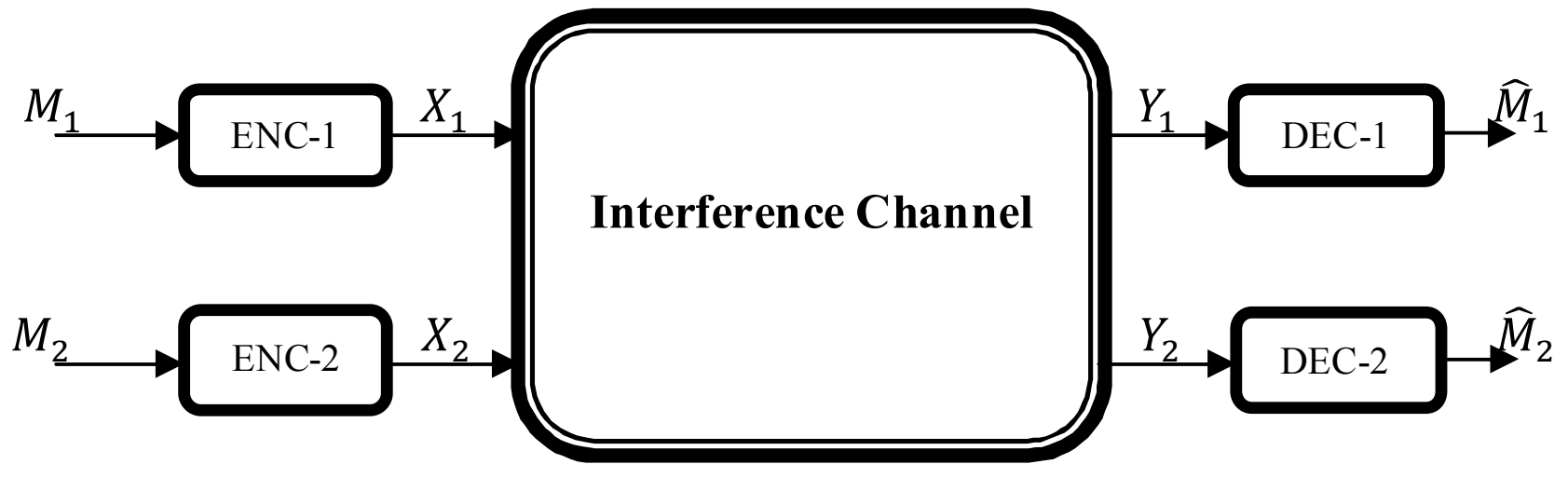

Figure 1. The two-user interference channel. The transmitter $X_i, i = 1,2$ sends the message $M_i$ over the channel.

The two-user IC is a communication scenario where two transmitters send independent messages to their respective receivers via a common media (see Fig. 1). The Gaussian fading channel is described by the following:

$$\begin{cases} Y_{1,t} = S_{11,t}X_{1,t} + S_{12,t}X_{2,t} + Z_{1,t} \\ Y_{2,t} = S_{21,t}X_{1,t} + S_{22,t}X_{2,t} + Z_{2,t} \end{cases}, \qquad t \geq 1 \tag{1}$$

The sequences $\{X_{1,t}\}_{t\geq1}$ and $\{X_{2,t}\}_{t\geq1}$ represent complex-valued transmitted signals by the transmitters, and $\{Y_{1,t}\}_{t\geq1}$ and $\{Y_{2,t}\}_{t\geq1}$ represent the received signals at the receivers. The sequences $\{Z_{1,t}\}_{t\geq1}$ and $\{Z_{2,t}\}_{t\geq1}$ denote additive noises each of which is an i.i.d. (complex) Gaussian random process with zero mean and unit variance. The state process of the channel is denoted by $\{\boldsymbol{S}_t = (S_{11,t}, S_{12,t}, S_{21,t}, S_{22,t})\}_{t\geq1}$ where the components $S_{11,t}, S_{12,t}, S_{21,t}$, and $S_{22,t}$ are complex-valued fading coefficients at the time instant $t$. We assume that the state process of the channel is a stationary and ergodic random process with limited energy which varies in time according to any arbitrary (known) probability distribution. The components of the state $\boldsymbol{S}_t = (S_{11,t}, S_{12,t}, S_{21,t}, S_{22,t}), t \geq 1$ are potentially correlated. Also, note that the state process of the channel is independent of the additive Gaussian noises.

As mentioned in introduction, we consider a scenario wherein both receivers perfectly know the state information while the transmitters have access to it partially. The partial side information at each transmitter is given by a deterministic function (potentially discrete-valued) of the channel state. Clearly, consider the following two deterministic functions:

$$\xi_i(.): \mathbb{C}^4 \to \mathcal{E}_i, \qquad i = 1,2$$

where $\mathcal{E}_i$ is an arbitrary (potentially finite) set. At each time instant $t, t \geq 1$, the transmitter $X_i, i = 1,2$ is informed of $E_{i,t} = \xi_i(\boldsymbol{S}_t)$ where $\boldsymbol{S}_t$ is the current state of the channel. The encoding and decoding schemes for the channel are described in details as follows.

*Encoding and decoding scheme:* For the two-user Gaussian fading IC (1), given a natural number $n$ and a pair $(R_1, R_2) \in \mathbb{R}_+^2$, a length- $n$ code $\mathfrak{C}^n(R_1, R_2)$ with two independent messages $M_1$ and $M_2$ uniformly distributed over the sets $\{1, \dots, 2^{nR_1}\}$ and $\{1, \dots, 2^{nR_2}\}$, respectively, consists of the following elements:

- ✓ Two sets of encoder functions $\{\mathfrak{E}_{1,t}\}_{t=1}^n$ and $\{\mathfrak{E}_{2,t}\}_{t=1}^n$ with:

$$\mathfrak{E}_{i,t}: \{1, \dots, 2^{nR_i}\} \times \mathcal{E}_i \to \mathbb{C}, \qquad i = 1,2$$

where the produced real-valued signals are as $X_{i,t} = \mathfrak{E}_{i,t}(M_i, E_{i,t}), t = 1, \dots, n$. The input signals are subjected to average power constraints as:

$$\frac{1}{n}\mathbb{E}\left[\sum_{t=1}^{n}\left|X_{i,t}\left(M_i, \xi_{i,t}(\boldsymbol{S}_t)\right)\right|^2\right] \leq P_i, \qquad i = 1,2$$

where $P_1$ and $P_2$ are arbitrary nonnegative real numbers.

- ✓ Two decoder functions $\mathfrak{D}_1(.)$ and $\mathfrak{D}_2(.)$ given as:

$$\mathfrak{D}_i: \mathbb{C}^n \times \mathbb{C}^{4n} \to \{1, \dots, 2^{nR_i}\}, \qquad i = 1,2$$

where the estimated messages are as $\widehat{M}_i = \mathfrak{D}_i(Y_i^n, \boldsymbol{S}^n)$.

The rate of the code is the pair $(R_1, R_2)$. Also, the average error probability of decoding is defined as:

$$P_e^{\mathfrak{C}^n} \triangleq Pr\left(\bigcup_{i=1,2}\{\mathfrak{D}_i(Y_i^n, \boldsymbol{S}^n) \neq M_i\}\right)$$

***Definition:*** For the two-user Gaussian fading IC (1), a pair $(R_1, R_2) \in \mathbb{R}_+^2$ is said to be achievable if there exist a sequence of codes $\mathfrak{C}^n(R_1, R_2)$ with $P_e^{\mathfrak{C}^n} \to 0$ as $n \to \infty$. The capacity region is the closure of the set of all achievable rates.

By these preliminaries, we are ready to state our main results.

# III. Main Results

In this section, we first propose a capacity inner bound for the Gaussian fading IC (1) with partial CSIT. This inner bound is derived based on the HK achievability scheme [20]. Next, we establish a novel outer bound on the capacity region. We also prove that the derived outer bound differs by no more than one bit per rate component from the inner bound if each transmitter has access to the interference to noise ratio perceived at its non-corresponding receiver. This yields the capacity region to within one bit. Finally, some special cases are presented where the given inner and outer bounds coincide and result to a full characterization of the capacity region.

Note that in the following analysis, the random variable $\boldsymbol{S} \triangleq (S_{11}, S_{12}, S_{21}, S_{22}) \in \mathbb{C}^4$ with a given known distribution $P_{\boldsymbol{S}}(\boldsymbol{s})$ represents the channel state. Also, $E_1 = \xi_1(\boldsymbol{S})$ and $E_2 = \xi_2(\boldsymbol{S})$ represent the partial side information at the transmitters $X_1$ and $X_2$, respectively.

***Proposition 1)*** *Define the rate region* $\mathfrak{R}_i^{GFIC}$ *as follows:*

$$\mathfrak{R}_i^{GFIC} \triangleq \bigcup_{\substack{\alpha(.):\mathcal{E}_1\to[0,1] \\ \beta(.):\mathcal{E}_2\to[0,1] \\ \varphi_i(.):\,\mathcal{E}_i\to\mathbb{R}_+,i=1,2 \\ \mathbb{E}[\varphi_i(E_i)]\le P_i,i=1,2}} \left\{\begin{array}{l}
(R_1,R_2)\in\mathbb{R}_+^2: \\
R_1 \le \mathbb{E}\left[\psi\left(\frac{|S_{11}|^2\varphi_1(E_1)}{|S_{12}|^2\varphi_2(E_2)\beta(E_2)+1}\right)\right] \\
R_2 \le \mathbb{E}\left[\psi\left(\frac{|S_{22}|^2\varphi_2(E_2)}{|S_{21}|^2\varphi_1(E_1)\alpha(E_1)+1}\right)\right] \\
R_1+R_2 \le \mathbb{E}\left[\psi\left(\frac{|S_{11}|^2\varphi_1(E_1)\alpha(E_1)}{|S_{12}|^2\varphi_2(E_2)\beta(E_2)+1}\right)\right] + \mathbb{E}\left[\psi\left(\frac{|S_{21}|^2\varphi_1(E_1)\big(1-\alpha(E_1)\big)+|S_{22}|^2\varphi_2(E_2)}{|S_{21}|^2\varphi_1(E_1)\alpha(E_1)+1}\right)\right], \\
R_1+R_2 \le \mathbb{E}\left[\psi\left(\frac{|S_{11}|^2\varphi_1(E_1)+|S_{12}|^2\varphi_2(E_2)\big(1-\beta(E_2)\big)}{|S_{12}|^2\varphi_2(E_2)\beta(E_2)+1}\right)\right] + \mathbb{E}\left[\psi\left(\frac{|S_{22}|^2\varphi_2(E_2)\beta(E_2)}{|S_{21}|^2\varphi_1(E_1)\alpha(E_1)+1}\right)\right], \\
R_1+R_2 \le \mathbb{E}\left[\psi\left(\frac{|S_{11}|^2\varphi_1(E_1)\alpha(E_1)+|S_{12}|^2\varphi_2(E_2)\big(1-\beta(E_2)\big)}{|S_{12}|^2\varphi_2(E_2)\beta(E_2)+1}\right)\right] \\
\qquad\qquad +\mathbb{E}\left[\psi\left(\frac{|S_{21}|^2\varphi_1(E_1)\big(1-\alpha(E_1)\big)+|S_{22}|^2\varphi_2(E_2)\beta(E_2)}{|S_{21}|^2\varphi_1(E_1)\alpha(E_1)+1}\right)\right], \\
2R_1+R_2 \le \mathbb{E}\left[\psi\left(\frac{|S_{11}|^2\varphi_1(E_1)+|S_{12}|^2\varphi_2(E_2)\big(1-\beta(E_2)\big)}{|S_{12}|^2\varphi_2(E_2)\beta(E_2)+1}\right)\right] + \mathbb{E}\left[\psi\left(\frac{|S_{11}|^2\varphi_1(E_1)\alpha(E_1)}{|S_{12}|^2\varphi_2(E_2)\beta(E_2)+1}\right)\right] \\
\qquad\qquad +\mathbb{E}\left[\psi\left(\frac{|S_{21}|^2\varphi_1(E_1)\big(1-\alpha(E_1)\big)+|S_{22}|^2\varphi_2(E_2)\beta(E_2)}{|S_{21}|^2\varphi_1(E_1)\alpha(E_1)+1}\right)\right], \\
R_1+2R_2 \le \mathbb{E}\left[\psi\left(\frac{|S_{22}|^2\varphi_2(E_2)\beta(E_2)}{|S_{21}|^2\varphi_1(E_1)\alpha(E_1)+1}\right)\right] + \mathbb{E}\left[\psi\left(\frac{|S_{21}|^2\varphi_1(E_1)\big(1-\alpha(E_1)\big)+|S_{22}|^2\varphi_2(E_2)}{|S_{21}|^2\varphi_1(E_1)\alpha(E_1)+1}\right)\right] \\
\qquad\qquad +\mathbb{E}\left[\psi\left(\frac{|S_{11}|^2\varphi_1(E_1)\alpha(E_1)+|S_{12}|^2\varphi_2(E_2)\big(1-\beta(E_2)\big)}{|S_{12}|^2\varphi_2(E_2)\beta(E_2)+1}\right)\right]
\end{array}\right\} \tag{2}$$

*The set* $\mathfrak{R}_i^{GFIC}$ *constitutes an inner bound on the capacity region of the two-user Gaussian fading IC in* (1).

*Proof of Proposition 1)* This achievable rate region is derived based on the HK message splitting scheme [20]. First note that the HK achievable rate region can be adapted for the Gaussian fading channel (1) as follows:

$$\bigcup_{P_Q P_{X_1W_1|E_1Q} P_{X_2W_2|E_2Q}} \left\{\begin{array}{l}
(R_1,R_2)\in\mathbb{R}_+^2: \\
R_1 \le I(X_1;Y_1|W_2,\boldsymbol{S},Q) \\
R_2 \le I(X_2;Y_2|W_1,\boldsymbol{S},Q) \\
R_1+R_2 \le I(X_1;Y_1|W_1,W_2,\boldsymbol{S},Q)+I(X_2,W_1;Y_2|\boldsymbol{S},Q) \\
R_1+R_2 \le I(X_2;Y_2|W_1,W_2,\boldsymbol{S},Q)+I(X_1,W_2;Y_1|\boldsymbol{S},Q) \\
R_1+R_2 \le I(X_1,W_2;Y_1|W_1,\boldsymbol{S},Q)+I(X_2,W_1;Y_2|W_2,\boldsymbol{S},Q) \\
2R_1+R_2 \le I(X_1,W_2;Y_1|\boldsymbol{S},Q)+I(X_1;Y_1|W_1,W_2,\boldsymbol{S},Q) \\
\qquad\qquad +I(X_2,W_1;Y_2|W_2,\boldsymbol{S},Q) \\
R_1+2R_2 \le I(X_2;Y_2|W_1,W_2,\boldsymbol{S},Q)+I(X_2,W_1;Y_2|\boldsymbol{S},Q) \\
\qquad\qquad +I(X_1,W_2;Y_1|W_1,\boldsymbol{S},Q)
\end{array}\right\} \tag{3}$$

where the power constraints on the transmitters are also considered as $\mathbb{E}[|X_i|^2] \leq P_i, i = 1,2$. Now, let $\varphi_i(.): \mathcal{E}_i \to \mathbb{R}_+$ be a power allocation policy function with respect to the transmitter $X_i$ where:

$$\mathbb{E}[\varphi_i(E_i)] \leq P_i, \qquad i = 1,2 \tag{4}$$

The input signals to achieve the region (2) are given as follows:

$$\begin{cases} X_1 \triangleq \sqrt{\varphi_1(E_1)}\left(\sqrt{\alpha(E_1)}V_1 + \sqrt{1-\alpha(E_1)}W_1\right) \\ X_2 \triangleq \sqrt{\varphi_2(E_2)}\left(\sqrt{\beta(E_2)}V_2 + \sqrt{1-\beta(E_2)}W_2\right) \end{cases} \tag{5}$$

where $V_1, V_2, W_1, W_2$ are independent complex Gaussian RVs with zero mean and unit variance; also, $\alpha(.): \mathcal{E}_1 \to [0,1]$ and $\beta(.): \mathcal{E}_2 \to [0,1]$ are two arbitrary deterministic functions. Note that this signaling in general represents a sub-optimal evaluation of the HK rate region (3) because it utilizes Gaussian input distributions and does not include time-sharing. In the HK scheme, each transmitter splits its message into two parts: a common part and a private part. The common parts of the messages are decoded at both receivers while the private parts are decoded only at their respective receivers (the receivers apply joint decoding technique), see [30] for details. According to the signaling scheme (5), the signals $V_1$ and $V_2$ are used to convey the private parts of the respective messages and $W_1$ and $W_2$ to convey the common parts. The functions $\alpha(.)$ and $\beta(.)$ determine the amount of power allocated to the common and private parts for the transmitter $X_1$ and $X_2$, respectively. Clearly, when the transmitters have access to side information (regardless of its quality) the portion of power allocated for transmission of each of the common and private messages naturally depends on the available side information. The functions $\alpha(.)$ and $\beta(.)$ in the signaling in (5) represent this fact. According to this signaling, the transmitter $X_1$ allocates the power $\varphi_1(E_1)\alpha(E_1)$ for transmission of the private message and the power $\varphi_1(E_1)\big(1-\alpha(E_1)\big)$ for transmission of the common message. Similarly, the transmitter $X_2$ allocates the power $\varphi_2(E_2)\beta(E_2)$ for transmission of the private message and the power $\varphi_2(E_2)\big(1-\beta(E_2)\big)$ for transmission of the common message. The derivation of the rate region (2) is complete. ■

***Remark 1:*** In the special case of $E_1 \equiv E_2 \equiv \boldsymbol{S}$, i.e., perfect state information at both transmitters, the signaling scheme in (5) was previously proposed in [12] for the channel.

As mentioned in introduction, one of the main contributions of this paper is to establish a novel outer bound on the capacity region of the channel. We derive our result in two steps. In the first step, we establish a single letter outer bound with constraints given by mutual information functions including some auxiliaries. A key feature in the derivation of this bound is novel applications of the Csiszar-Korner identity [31]. This result is presented in the following Lemma.

***Lemma 1)*** *Define the rate region $\mathfrak{R}_o^{UV\to GFIC}$ as follows:*

$$\mathfrak{R}_o^{UV\to GFIC} \triangleq \bigcup_{\substack{P_Q P_{X_1|E_1Q} P_{X_2|E_2Q} \\ \times P_{UV|X_1X_2SQ} \\ , \\ \mathbb{E}[|X_i|^2]\leq P_i, i=1,2}} \left\{ \begin{array}{l} (R_1,R_2) \in \mathbb{R}_+^2 : \\ G_1 \triangleq S_{21}X_1 + Z_2, \quad G_2 \triangleq S_{12}X_2 + Z_1 \\ R_1 \leq \min\left\{ \begin{array}{c} I(X_1;Y_1|X_2,\boldsymbol{S},Q), \\ I(U,X_1;Y_1|\boldsymbol{S},Q) \end{array} \right\}, \\ R_2 \leq \min\left\{ \begin{array}{c} I(X_2;Y_2|X_1,\boldsymbol{S},Q), \\ I(V,X_2;Y_2|\boldsymbol{S},Q) \end{array} \right\}, \\ R_1 + R_2 \leq I(X_1;Y_1|V,X_2,\boldsymbol{S},Q) + I(V,X_2;Y_2|\boldsymbol{S},Q) \\ R_1 + R_2 \leq I(X_2;Y_2|U,X_1,\boldsymbol{S},Q) + I(U,X_1;Y_1|\boldsymbol{S},Q) \\ R_1 + R_2 \leq I(X_1,X_2;Y_1|G_1,\boldsymbol{S},Q) + I(X_1,X_2;Y_2|G_2,\boldsymbol{S},Q) \\ , \\ 2R_1 + R_2 \leq I(X_1;Y_1|V,X_2,\boldsymbol{S},Q) + I(V,X_2;Y_2|G_2,\boldsymbol{S},Q) \\ \qquad\qquad + I(X_1,X_2;Y_1|\boldsymbol{S},Q) \\ R_1 + 2R_2 \leq I(X_2;Y_2|U,X_1,\boldsymbol{S},Q) + I(U,X_1;Y_1|G_1,\boldsymbol{S},Q) \\ \qquad\qquad + I(X_1,X_2;Y_2|\boldsymbol{S},Q) \end{array} \right\} \tag{6}$$

*The set* $\mathfrak{R}_o^{UV\to GFIC}$ *constitutes an outer bound on the capacity region of the two-user Gaussian fading IC in (1).*

*Proof of Lemma 1)* Consider a length-$n$ code with the rate $(R_1,R_2)$ and vanishing average error probability for the channel. Based on the Fano's inequality we have:

$$H(M_1|Y_1^n,\boldsymbol{S}^n)\leq n\epsilon_{1,n}$$
$$H(M_1|Y_2^n,\boldsymbol{S}^n)\leq n\epsilon_{2,n}$$

where $\epsilon_{1,n},\epsilon_{2,n}\to 0$ as $n\to\infty$. Define new auxiliary random variables $U_t$ and $V_t$ as follows:

$$\begin{cases}U_t\triangleq\left(Y_{1,t+1}^n,Y_2^{t-1},M_1,\boldsymbol{S}^{t-1},\boldsymbol{S}_{t+1}^n\right)\\ V_t\triangleq\left(Y_{1,t+1}^n,Y_2^{t-1},M_2,\boldsymbol{S}^{t-1},\boldsymbol{S}_{t+1}^n\right)\end{cases},\qquad t=1,\dots,n \tag{7}$$

Note that $X_{1,t}$ and $X_{2,t}$ are given by deterministic functions of $\left(U_t,E_{1,t}=\xi_1(\boldsymbol{S}_t)\right)$ and $\left(V_t,E_{2,t}=\xi_2(\boldsymbol{S}_t)\right)$, respectively. Moreover, define the genie signals $G_{1,t}$ and $G_{2,t}$ as:

$$\begin{cases}G_{1,t}\triangleq S_{21,t}X_{1,t}+Z_{2,t}\\ G_{2,t}\triangleq S_{12,t}X_{2,t}+Z_{1,t}\end{cases},\qquad t=1,\dots,n \tag{8}$$

Now, one can write:

$$nR_1\overset{(a)}{\leq}\sum_{t=1}^n I\left(X_{1,t};Y_{1,t}\middle|X_{2,t},\boldsymbol{S}_t\right)+n\epsilon_{1,n} \tag{9}$$

where inequality (a) is a standard cut-set bound. Also,

$$\begin{aligned}nR_1&\leq I(M_1;Y_1^n,\boldsymbol{S}^n)+n\epsilon_{1,n}\overset{(a)}{=}I(M_1;Y_1^n|\boldsymbol{S}^n)+n\epsilon_{1,n}\\
&=\sum_{t=1}^n I\left(M_1;Y_{1,t}\middle|Y_{1,t+1}^n,\boldsymbol{S}^n\right)+n\epsilon_{1,n}\\
&\leq\sum_{t=1}^n I\left(Y_{1,t+1}^n,Y_2^{t-1},M_1,\boldsymbol{S}^{t-1},\boldsymbol{S}_{t+1}^n;Y_{1,t}\middle|\boldsymbol{S}_t\right)+n\epsilon_{1,n}\\
&=\sum_{t=1}^n I\left(U_t;Y_{1,t}\middle|\boldsymbol{S}_t\right)+n\epsilon_{1,n}=\sum_{t=1}^n I\left(U_t,X_{1,t};Y_{1,t}\middle|\boldsymbol{S}_t\right)+n\epsilon_{1,n}\end{aligned} \tag{10}$$

where equality (a) holds because $M_1$ and $\boldsymbol{S}^n$ are independent. For the sum-rate, we can write:

$$\begin{aligned}n(R_1+R_2)&\leq I(M_2;Y_2^n|M_1,\boldsymbol{S}^n)+I(M_1;Y_1^n,\boldsymbol{S}^n)+n\left(\epsilon_{1,n}+\epsilon_{2,n}\right)\\
&=\sum_{t=1}^n I\left(M_2;Y_{2,t}\middle|Y_2^{t-1},M_1,\boldsymbol{S}^n\right)+\sum_{t=1}^n I\left(M_1;Y_{1,t}\middle|Y_{1,t+1}^n,\boldsymbol{S}^n\right)+n\left(\epsilon_{1,n}+\epsilon_{2,n}\right)\\
&\leq\sum_{t=1}^n I\left(M_2,Y_{1,t+1}^n;Y_{2,t}\middle|Y_2^{t-1},M_1,\boldsymbol{S}^n\right)+\sum_{t=1}^n I\left(Y_2^{t-1},M_1;Y_{1,t}\middle|Y_{1,t+1}^n,\boldsymbol{S}^n\right)\\
&\qquad-\sum_{t=1}^n I\left(Y_2^{t-1};Y_{1,t}\middle|Y_{1,t+1}^n,M_1,\boldsymbol{S}^n\right)+n\left(\epsilon_{1,n}+\epsilon_{2,n}\right)\\
&=\sum_{t=1}^n I\left(M_2;Y_{2,t}\middle|Y_{1,t+1}^n,Y_2^{t-1},M_1,\boldsymbol{S}^n\right)+\sum_{t=1}^n I\left(Y_2^{t-1},M_1;Y_{1,t}\middle|Y_{1,t+1}^n,\boldsymbol{S}^n\right)\\
&\qquad+\sum_{t=1}^n I\left(Y_{1,t+1}^n;Y_{2,t}\middle|Y_2^{t-1},M_1,\boldsymbol{S}^n\right)-\sum_{t=1}^n I\left(Y_2^{t-1};Y_{1,t}\middle|Y_{1,t+1}^n,M_1,\boldsymbol{S}^n\right)+n\left(\epsilon_{1,n}+\epsilon_{2,n}\right)\\
&\overset{(a)}{=}\sum_{t=1}^n I\left(M_2;Y_{2,t}\middle|Y_{1,t+1}^n,Y_2^{t-1},M_1,\boldsymbol{S}^n\right)+\sum_{t=1}^n I\left(Y_2^{t-1},M_1;Y_{1,t}\middle|Y_{1,t+1}^n,\boldsymbol{S}^n\right)+n\left(\epsilon_{1,n}+\epsilon_{2,n}\right)\\
&\overset{(b)}{\leq}\sum_{t=1}^n I\left(M_2;Y_{2,t}\middle|Y_{1,t+1}^n,Y_2^{t-1},M_1,\boldsymbol{S}^{t-1},\boldsymbol{S}_{t+1}^n,\boldsymbol{S}_t\right)+\sum_{t=1}^n I\left(Y_{1,t+1}^n,Y_2^{t-1},M_1,\boldsymbol{S}^{t-1},\boldsymbol{S}_{t+1}^n;Y_{1,t}\middle|\boldsymbol{S}_t\right)+n\left(\epsilon_{1,n}+\epsilon_{2,n}\right)\\
&\overset{(c)}{=}\sum_{t=1}^n I\left(X_{2,t},M_2;Y_{2,t}\middle|U_t,X_{1,t},\boldsymbol{S}_t\right)+\sum_{t=1}^n I\left(U_t,X_{1,t};Y_{1,t}\middle|\boldsymbol{S}_t\right)+n\left(\epsilon_{1,n}+\epsilon_{2,n}\right)\\
&\overset{(d)}{=}\sum_{t=1}^n I\left(X_{2,t};Y_{2,t}\middle|U_t,X_{1,t},\boldsymbol{S}_t\right)+\sum_{t=1}^n I\left(U_t,X_{1,t};Y_{1,t}\middle|\boldsymbol{S}_t\right)+n\left(\epsilon_{1,n}+\epsilon_{2,n}\right)\end{aligned} \tag{11}$$

where equality (a) is due to the Csiszar-Korner identity, inequality (b) holds because conditioning does not increase the entropy, equality (c) holds because $X_{2,t}$ is a deterministic function of $\left(M_2, E_{2,t} = \xi_2(\boldsymbol{S}_t)\right)$, and equality (d) holds because $M_2, U_t \to X_{1,t}, X_{2,t}, \boldsymbol{S}_t \to Y_{2,t}$ forms a Markov chain. Next, we bound the linear combination $R_1 + 2R_2$;

$$
\begin{aligned}
n(R_1+2R_2) &\le I(M_2;Y_2^n,\boldsymbol{S}^n) + I(M_2;Y_2^n,M_1,\boldsymbol{S}^n) + I(M_1;Y_1^n,G_1^n,\boldsymbol{S}^n) + n(\epsilon_{1,n}+2\epsilon_{2,n}) \\
&= I(X_2^n;Y_2^n|\boldsymbol{S}^n) + I(M_2;Y_2^n|M_1,\boldsymbol{S}^n) + I(M_1;Y_1^n|G_1^n,\boldsymbol{S}^n) + I(X_1^n;G_1^n|\boldsymbol{S}^n) + n(\epsilon_{1,n}+2\epsilon_{2,n}) \\
&= H(Y_2^n|\boldsymbol{S}^n) - H(Y_2^n|X_2^n,\boldsymbol{S}^n) + I(M_2;Y_2^n|M_1,\boldsymbol{S}^n) + H(Y_1^n|G_1^n,\boldsymbol{S}^n) - H(Y_1^n|M_1,\boldsymbol{S}^n) + H(G_1^n|\boldsymbol{S}^n) - H(Z_2^n) \\
&\qquad + n(\epsilon_{1,n}+2\epsilon_{2,n}) \\
&\overset{(a)}{=} H(Y_2^n|\boldsymbol{S}^n) + I(M_2;Y_2^n|M_1,\boldsymbol{S}^n) + I(M_1;Y_1^n|\boldsymbol{S}^n) - H(Y_1^n|\boldsymbol{S}^n) + H(Y_1^n|G_1^n,\boldsymbol{S}^n) - H(Z_2^n) + n(\epsilon_{1,n}+2\epsilon_{2,n}) \\
&\overset{(b)}{\le} H(Y_2^n|\boldsymbol{S}^n) + \textstyle\sum_{t=1}^n I\left(M_2;Y_{2,t}\middle|Y_{1,t+1}^n,Y_2^{t-1},M_1,\boldsymbol{S}^n\right) + \sum_{t=1}^n I\left(Y_2^{t-1},M_1;Y_{1,t}\middle|Y_{1,t+1}^n,\boldsymbol{S}^n\right) \\
&\qquad - H(Y_1^n|\boldsymbol{S}^n) + H(Y_1^n|G_1^n,\boldsymbol{S}^n) - H(Z_2^n) + n(\epsilon_{1,n}+2\epsilon_{2,n}) \\
&= H(Y_2^n|\boldsymbol{S}^n) + \textstyle\sum_{t=1}^n I\left(M_2;Y_{2,t}\middle|Y_{1,t+1}^n,Y_2^{t-1},M_1,\boldsymbol{S}^{t-1},\boldsymbol{S}_{t+1}^n,\boldsymbol{S}_t\right) + \sum_{t=1}^n I\left(Y_{1,t+1}^n,Y_2^{t-1},M_1,\boldsymbol{S}^{t-1},\boldsymbol{S}_{t+1}^n;Y_{1,t}\middle|\boldsymbol{S}_t\right) \\
&\qquad - \textstyle\sum_{t=1}^n I\left(Y_{1,t+1}^n,\boldsymbol{S}^{t-1},\boldsymbol{S}_{t+1}^n;Y_{1,t}\middle|\boldsymbol{S}_t\right) - H(Y_1^n|\boldsymbol{S}^n) + H(Y_1^n|G_1^n,\boldsymbol{S}^n) - H(Z_2^n) + n(\epsilon_{1,n}+2\epsilon_{2,n}) \\
&= H(Y_2^n|\boldsymbol{S}^n) + \textstyle\sum_{t=1}^n I\left(X_{2,t};Y_{2,t}\middle|U_t,X_{1,t},\boldsymbol{S}_t\right) + \sum_{t=1}^n I\left(U_t,X_{1,t};Y_{1,t}\middle|\boldsymbol{S}_t\right) \\
&\qquad - \textstyle\sum_{t=1}^n H\left(Y_{1,t}\middle|\boldsymbol{S}_t\right) + \sum_{t=1}^n H\left(Y_{1,t}\middle|Y_{1,t+1}^n,\boldsymbol{S}^n\right) - H(Y_1^n|\boldsymbol{S}^n) + H(Y_1^n|G_1^n,\boldsymbol{S}^n) - H(Z_2^n) + n(\epsilon_{1,n}+2\epsilon_{2,n}) \\
&= H(Y_2^n|\boldsymbol{S}^n) + \textstyle\sum_{t=1}^n I\left(X_{2,t};Y_{2,t}\middle|U_t,X_{1,t},\boldsymbol{S}_t\right) + \sum_{t=1}^n I\left(U_t,X_{1,t};Y_{1,t}\middle|\boldsymbol{S}_t\right) - \sum_{t=1}^n H\left(Y_{1,t}\middle|\boldsymbol{S}_t\right) + H(Y_1^n|G_1^n,\boldsymbol{S}^n) - H(Z_2^n) \\
&\qquad + n(\epsilon_{1,n}+2\epsilon_{2,n}) \\
&\overset{(c)}{=} \textstyle\sum_{t=1}^n I\left(X_{2,t};Y_{2,t}\middle|U_t,X_{1,t},\boldsymbol{S}_t\right) - \sum_{t=1}^n H\left(Y_{1,t}\middle|U_t,X_{1,t},\boldsymbol{S}_t\right) + H(Y_1^n|G_1^n,\boldsymbol{S}^n) + I(X_1^n,X_2^n;Y_2^n|\boldsymbol{S}^n) + n(\epsilon_{1,n}+2\epsilon_{2,n}) \\
&\le \textstyle\sum_{t=1}^n I\left(X_{2,t};Y_{2,t}\middle|U_t,X_{1,t},\boldsymbol{S}_t\right) - \sum_{t=1}^n H\left(Y_{1,t}\middle|U_t,X_{1,t},\boldsymbol{S}_t\right) + \sum_{t=1}^n H\left(Y_{1,t}\middle|G_{1,t},\boldsymbol{S}_t\right) + \sum_{t=1}^n I\left(X_{1,t},X_{2,t};Y_{2,t}\middle|\boldsymbol{S}_t\right) \\
&\qquad + n(\epsilon_{1,n}+2\epsilon_{2,n}) \\
&\overset{(d)}{=} \textstyle\sum_{t=1}^n I\left(X_{2,t};Y_{2,t}\middle|U_t,X_{1,t},\boldsymbol{S}_t\right) + \sum_{t=1}^n I\left(U_t,X_{1,t};Y_{1,t}\middle|G_{1,t},\boldsymbol{S}_t\right) + \sum_{t=1}^n I\left(X_{1,t},X_{2,t};Y_{2,t}\middle|\boldsymbol{S}_t\right) + n(\epsilon_{1,n}+2\epsilon_{2,n})
\end{aligned}
\tag{12}
$$

where equality (a) holds because $H(Y_2^n|X_2^n,\boldsymbol{S}^n) = H(G_1^n|\boldsymbol{S}^n)$, inequality (b) is derived similar to the equation (a) of (11), equality (c) holds because $H(Y_2^n|X_1^n,X_2^n,\boldsymbol{S}^n) = H(Z_2^n)$, and equability (d) holds because given $(X_{1,t},\boldsymbol{S}_t)$, the genie signal $G_{1,t}$ is reduced to $Z_{2,t}$ which is independent of $U_t, X_{1,t}, Y_{1,t}$, and $\boldsymbol{S}_t$.

By following a rather similar procedure, one can derive:

$$
\begin{aligned}
nR_2 &\le \textstyle\sum_{t=1}^n I\left(X_{2,t};Y_{2,t}\middle|X_{1,t},\boldsymbol{S}_t\right) + n\epsilon_{2,n} \\
nR_2 &\le \textstyle\sum_{t=1}^n I\left(V_t,X_{2,t};Y_{2,t}\middle|\boldsymbol{S}_t\right) + n\epsilon_{2,n} \\
n(R_1+R_2) &\le \textstyle\sum_{t=1}^n I\left(X_{1,t};Y_{1,t}\middle|V_t,X_{2,t},\boldsymbol{S}_t\right) + \sum_{t=1}^n I\left(V_t,X_{2,t};Y_{2,t}\middle|\boldsymbol{S}_t\right) + n(\epsilon_{1,n}+\epsilon_{2,n}) \\
n(2R_1+R_2) &\le \textstyle\sum_{t=1}^n I\left(X_{1,t};Y_{1,t}\middle|V_t,X_{2,t},\boldsymbol{S}_t\right) + \sum_{t=1}^n I\left(V_t,X_{2,t};Y_{2,t}\middle|G_{2,t},\boldsymbol{S}_t\right) + I\left(X_{1,t},X_{2,t};Y_{1,t}\middle|\boldsymbol{S},Q\right) + n(\epsilon_{1,n}+\epsilon_{2,n})
\end{aligned}
\tag{13}
$$

Also, we can obtain the following bound on the sum-rate:

$$
\begin{aligned}
n(R_1+R_2) &\le I(X_1^n;Y_1^n,G_1^n|\boldsymbol{S}^n) + I(X_2^n;Y_2^n,G_2^n|\boldsymbol{S}^n) + n(\epsilon_{1,n}+\epsilon_{2,n}) \\
&= I(X_1^n;G_1^n|\boldsymbol{S}^n) + I(X_1^n;Y_1^n|G_1^n,\boldsymbol{S}^n) + I(X_2^n;G_2^n|\boldsymbol{S}^n) + I(X_2^n;Y_2^n|G_2^n,\boldsymbol{S}^n) + n(\epsilon_{1,n}+\epsilon_{2,n}) \\
&\overset{(a)}{=} H(G_1^n|\boldsymbol{S}^n) - H(G_1^n|X_1^n,\boldsymbol{S}^n) + H(Y_1^n|G_1^n,\boldsymbol{S}^n) - H(Y_1^n|X_1^n,\boldsymbol{S}^n)
\end{aligned}
$$

$$
\begin{aligned}
&\quad +H(G_2^n|\boldsymbol{S}^n) - H(G_2^n|X_2^n,\boldsymbol{S}^n) + H(Y_2^n|G_2^n,\boldsymbol{S}^n) - H(Y_2^n|X_2^n,\boldsymbol{S}^n) + n(\epsilon_{1,n}+\epsilon_{2,n})\\
&= H(G_1^n|\boldsymbol{S}^n) - H(Z_2^n) + H(Y_1^n|G_1^n,\boldsymbol{S}^n) - H(G_2^n|\boldsymbol{S}^n)\\
&\quad +H(G_2^n|\boldsymbol{S}^n) - H(Z_1^n) + H(Y_2^n|G_2^n,\boldsymbol{S}^n) - H(G_1^n|\boldsymbol{S}^n) + n(\epsilon_{1,n}+\epsilon_{2,n})\\
&\overset{(b)}{=} H(Y_1^n|G_1^n,\boldsymbol{S}^n) - H(Y_1^n|X_1^n,X_2^n,\boldsymbol{S}^n) + H(Y_2^n|G_2^n,\boldsymbol{S}^n) - H(Y_2^n|X_1^n,X_2^n,\boldsymbol{S}^n) + n(\epsilon_{1,n}+\epsilon_{2,n})\\
&= H(Y_1^n|G_1^n,\boldsymbol{S}^n) - H(Y_1^n|X_1^n,X_2^n,G_1^n,\boldsymbol{S}^n) + H(Y_2^n|G_2^n,\boldsymbol{S}^n) - H(Y_2^n|X_1^n,X_2^n,G_2^n,\boldsymbol{S}^n) + n(\epsilon_{1,n}+\epsilon_{2,n})\\
&= I(X_1^n,X_2^n;Y_1^n|G_1^n,\boldsymbol{S}^n) + I(X_1^n,X_2^n;Y_2^n|G_2^n,\boldsymbol{S}^n) + n(\epsilon_{1,n}+\epsilon_{2,n})\\
&\leq \sum_{t=1}^n I(X_{1,t},X_{2,t};Y_{1,t}|G_{1,t},\boldsymbol{S}_t) + \sum_{t=1}^n I(X_{1,t},X_{2,t};Y_{2,t}|G_{2,t},\boldsymbol{S}_t) + n(\epsilon_{1,n}+\epsilon_{2,n})
\end{aligned}
\tag{14}
$$

where equality (a) holds because $G_i^n \to X_i^n,\boldsymbol{S}^n \to Y_i^n, i=1,2$ forms a Markov chain, and equality (b) holds because $H(Y_i^n|X_1^n,X_2^n,\boldsymbol{S}^n) = H(Z_i^n), i=1,2$.

We next introduce a time-sharing random variable $Q$ uniformly distributed over the set $\{1,\ldots,n\}$ and independent of other RVs. Define:

$$
\begin{cases}
\boldsymbol{S} \triangleq \boldsymbol{S}_Q & \\
U \triangleq U_Q, & V \triangleq V_Q\\
X_1 \triangleq X_{1,Q}, & X_2 \triangleq X_{2,Q}\\
Y_1 \triangleq Y_{1,Q}, & Y_2 \triangleq Y_{2,Q}\\
Z_1 \triangleq Z_{1,Q}, & Z_2 \triangleq Z_{2,Q}\\
G_1 \triangleq G_{1,Q}, & G_2 \triangleq G_{2,Q}
\end{cases}
\tag{15}
$$

We remark that $\boldsymbol{S}, Z_1$, and $Z_2$ are independent of $Q$ because the state process of the channel is stationary and the additive noises are i.i.d. Now, based on (15) we can write:

$$
\begin{aligned}
R_1 \leq \sum_{t=1}^n \frac{1}{n} I(X_{1,t};Y_{1,t}|X_{2,t},\boldsymbol{S}_t) + \epsilon_{1,n} &= I(X_{1,Q};Y_{1,Q}|X_{2,Q},\boldsymbol{S}_Q,Q) + \epsilon_{1,n}\\
&= I(X_1;Y_1|X_2,\boldsymbol{S},Q) + \epsilon_{1,n}
\end{aligned}
\tag{16}
$$

Other summations included in the bounds (10)-(14) can also be expressed in such a compact form, similarly. Finally, by letting $n \to \infty$, we obtain the constraints in (6). Note that the input signals $X_1$ and $X_2$ must satisfy the following:

$$
\mathbb{E}[|X_i|^2] = \frac{1}{n}\mathbb{E}\left[\sum_{t=1}^n |X_{i,t}|^2\right] \leq P_i, \qquad i=1,2
\tag{17}
$$

The proof of Lemma 1 is thus complete. ■

***Remarks 2:***

1. Note that in the outer bound $\mathfrak{R}_o^{UV\to GFIC}$ in (6) since the state process is stationary, both auxiliary random variables $U$ and $V$ are correlated with the state random variable $\boldsymbol{S}$. Nevertheless, if we impose that the state process of the channel is memoryless (i.i.d. over time), they are both independent of $\boldsymbol{S}$.
2. The first four constraints of (6) are actually adapted from the UV-outer bound for the discrete IC[2] established in our recent paper [2, Sec. III.B]. We refer to the region $\mathfrak{R}_o^{UV\to GFIC}$ in (6) as the UV-outer bound for the Gaussian fading IC. Note that it can be directly reduced to an outer bound for the Gaussian IC with fixed channel gains, as well.
3. Consider the novel constraints given in (6) on the linear combinations $2R_1+R_2$ and $R_1+2R_2$. These constraints are characterized based on both the genie signals and the auxiliaries $U$ and $V$. The inclusion of the auxiliaries $U$ and $V$ into these constraints (that is derived by a subtle application of the Csiszar-Korner identity) puts them in correlation with the constraints on the sum-rate and the individual rates. As will be shown, this new technique yields tighter outer bounds than the arguments based on the worst case noise lemma [27].

[2] We argue later that these constraints are actually an alternative characterization of the so-called broadcast channel outer bound [23] (or the Kramer outer bound) for the Gaussian IC. The main advantage of this new characterization is that it can be simply adapted for the fading channel, as given in (6).

4. As we see later, the UV-outer bound (6) simultaneously includes several benefits: 1) It is optimal for the channel with strong interference, 2) It is sum-rate optimal for the channel with mixed interference, 3) It is to within one bit of the capacity region for all range of channel parameters, 4) It has a single-letter characterization which is not as complex as the ones derived in [21-26]. Moreover, it is strictly tighter than the Kramer's outer bounds [32] and the ETW outer bounds [7] for all range of channel parameters where the capacity is unknown. In general, the characterization (6) exhibits an efficient combination of the broadcast channel techniques (i.e., manipulating mutual information functions composed of vector RVs by Csiszar-Korner identity) and the genie-aided techniques for establishing capacity outer bounds.

In the second step, by evaluating the constraints of the UV-outer bound (6) we derive a bound with an explicit characterization. A remarkable point is that we only make use of the EPI for this purpose. In fact, we essentially follow a similar approach to the one developed in our concurrent paper [29] for evaluating the UV-outer bound for the two-user fading broadcast channel. This derivation is given in the following theorem.

***Theorem 1)*** *Consider the two-user Gaussian fading IC in (1). Define the rate region* $\mathfrak{R}_o^{GFIC}$ *as follows:*

$$\mathfrak{R}_o^{GFIC} \triangleq \bigcup_{\substack{\alpha(.),\beta(.)\\ \varphi_1(.),\varphi_2(.)}} \left\{\begin{array}{l}
(R_1,R_2)\in\mathbb{R}_+^2 : \\
R_1 \leq \min\left\{\begin{array}{l}\mathbb{E}\left[\psi\left(|S_{11}|^2\varphi_1(E_1)\right)\right],\\ \mathbb{E}\left[\psi\left(\dfrac{|S_{11}|^2\varphi_1(E_1)+|S_{12}|^2\left(1-\alpha(\boldsymbol{S})\right)\varphi_2(E_2)}{|S_{12}|^2\alpha(\boldsymbol{S})\varphi_2(E_2)+1}\right)\mathbb{1}(|S_{12}|<|S_{22}|)\right]\\ \qquad +\mathbb{E}\left[\psi\left(|S_{11}|^2\varphi_1(E_1)+|S_{12}|^2\varphi_2(E_2)\right)\mathbb{1}(|S_{12}|\geq|S_{22}|)\right]\end{array}\right\},\\
R_2 \leq \min\left\{\begin{array}{l}\mathbb{E}\left[\psi\left(S_{22}^2\varphi_2(E_2)\right)\right],\\ \mathbb{E}\left[\psi\left(\dfrac{|S_{21}|^2\left(1-\beta(\boldsymbol{S})\right)\varphi_1(E_1)+|S_{22}|^2\varphi_2(E_2)}{|S_{21}|^2\beta(\boldsymbol{S})\varphi_1(E_1)+1}\right)\mathbb{1}(|S_{21}|<|S_{11}|)\right]\\ \qquad +\mathbb{E}\left[\psi\left(|S_{21}|^2\varphi_1(E_1)+|S_{22}|^2\varphi_2(E_2)\right)\mathbb{1}(|S_{21}|\geq|S_{11}|)\right]\end{array}\right\},\\
R_1+R_2 \leq \mathbb{E}\left[\left(\psi\left(|S_{11}|^2\beta(\boldsymbol{S})\varphi_1(E_1)\right)+\psi\left(\dfrac{|S_{21}|^2\left(1-\beta(\boldsymbol{S})\right)\varphi_1(E_1)+|S_{22}|^2\varphi_2(E_2)}{|S_{21}|^2\beta(\boldsymbol{S})\varphi_1(E_1)+1}\right)\right)\mathbb{1}(|S_{21}|<|S_{11}|)\right]\\
\qquad\qquad +\mathbb{E}\left[\psi\left(|S_{21}|^2\varphi_1(E_1)+|S_{22}|^2\varphi_2(E_2)\right)\mathbb{1}(|S_{21}|\geq|S_{11}|)\right],\\
R_1+R_2 \leq \mathbb{E}\left[\left(\psi\left(|S_{22}|^2\alpha(\boldsymbol{S})\varphi_2(E_2)\right)+\psi\left(\dfrac{|S_{11}|^2\varphi_1(E_1)+|S_{12}|^2\left(1-\alpha(\boldsymbol{S})\right)\varphi_2(E_2)}{|S_{12}|^2\alpha(\boldsymbol{S})\varphi_2(E_2)+1}\right)\right)\mathbb{1}(|S_{12}|<|S_{22}|)\right]\\
\qquad\qquad +\mathbb{E}\left[\psi\left(|S_{11}|^2\varphi_1(E_1)+|S_{12}|^2\varphi_2(E_2)\right)\mathbb{1}(|S_{12}|\geq|S_{22}|)\right],\\
R_1+R_2 \leq \mathbb{E}\left[\psi\left(|S_{12}|^2\varphi_2(E_2)+\dfrac{|S_{11}|^2\varphi_1(E_1)}{|S_{21}|^2\varphi_1(E_1)+1}\right)\right]+\mathbb{E}\left[\psi\left(|S_{21}|^2\varphi_1(E_1)+\dfrac{|S_{22}|^2\varphi_2(E_2)}{|S_{12}|^2\varphi_2(E_2)+1}\right)\right],\\
2R_1+R_2 \leq \mathbb{E}\left[\left(\psi\left(|S_{11}|^2\beta(\boldsymbol{S})\varphi_1(E_1)\right)-\psi\left(|S_{21}|^2\beta(\boldsymbol{S})\varphi_1(E_1)\right)\right)\mathbb{1}(|S_{21}|<|S_{11}|)\right]\\
\qquad\qquad +\mathbb{E}\left[\psi\left(|S_{21}|^2\varphi_1(E_1)+\dfrac{|S_{22}|^2\varphi_2(E_2)}{|S_{12}|^2\varphi_2(E_2)+1}\right)\right]+\mathbb{E}\left[\psi\left(|S_{11}|^2\varphi_1(E_1)+|S_{12}|^2\varphi_2(E_2)\right)\right],\\
R_1+2R_2 \leq \mathbb{E}\left[\left(\psi\left(|S_{22}|^2\alpha(\boldsymbol{S})\varphi_2(E_2)\right)-\psi\left(|S_{12}|^2\alpha(\boldsymbol{S})\varphi_2(E_2)\right)\right)\mathbb{1}(|S_{12}|<|S_{22}|)\right]\\
\qquad\qquad +\mathbb{E}\left[\psi\left(|S_{12}|^2\varphi_2(E_2)+\dfrac{|S_{11}|^2\varphi_1(E_1)}{|S_{21}|^2\varphi_1(E_1)+1}\right)\right]+\mathbb{E}\left[\psi\left(|S_{21}|^2\varphi_1(E_1)+|S_{22}|^2\varphi_2(E_2)\right)\right]
\end{array}\right\} \tag{18}$$

*where* $\alpha(.):\mathbb{C}^4\to[0,1]$ *and* $\beta(.):\mathbb{C}^4\to[0,1]$ *are arbitrary deterministic functions; also,* $\varphi_i(.):\mathcal{E}_i\to\mathbb{R}_+$ *with* $\mathbb{E}[\varphi_i(E_i)]\leq P_i$ *denotes the power allocation policy for the transmitter* $X_i$, $i=1,2$. *The set* $\mathfrak{R}_o^{GFIC}$ *constitutes an outer bound on the capacity region.*

*Proof of Theorem 1)* We derive this outer bound by evaluating the UV-outer bound in (6). We only evaluate the following constraints:

$$\begin{aligned} R_1 &\leq I(X_1;Y_1|X_2,\boldsymbol{S},Q) && \text{(a)}\\ R_1 &\leq I(U,X_1;Y_1|\boldsymbol{S},Q) && \text{(b)}\\ R_1+R_2 &\leq I(X_2;Y_2|U,X_1,\boldsymbol{S},Q)+I(U,X_1;Y_1|\boldsymbol{S},Q) && \text{(c)}\\ R_1+2R_2 &\leq I(X_2;Y_2|U,X_1,\boldsymbol{S},Q)+I(U,X_1;Y_1|G_1,\boldsymbol{S},Q)+I(X_1,X_2;Y_2|\boldsymbol{S},Q) && \text{(d)}\\ R_1+R_2 &\leq I(X_1,X_2;Y_1|G_1,\boldsymbol{S},Q)+I(X_1,X_2;Y_2|G_2,\boldsymbol{S},Q) && \text{(e)} \end{aligned} \tag{19}$$

The other constraints are evaluated symmetrically. Fix a joint PDF $P_Q P_{X_1|E_1Q} P_{X_2|E_2Q} P_{UV|X_1X_2SQ}$ with $\mathbb{E}[|X_i|^2] \leq P_i,\ i=1,2$. Define the deterministic functions $\varphi_1(.)$ and $\varphi_2(.)$ as follows:

$$\varphi_i(.):\mathcal{E}_i \to \mathbb{R}_+, \qquad \varphi_i(e_i) \triangleq \mathbb{E}[|X_i|^2|E_i=e_i], \qquad i=1,2 \tag{20}$$

Thereby, we have: $\mathbb{E}[\varphi_i(E_i)] \leq P_i$. First let us evaluate the constraint (19, a). We can write:

$$\begin{aligned} R_1 \leq I(X_1;Y_1|X_2,\boldsymbol{S},Q) &= H(S_{11}X_1+S_{12}X_2+Z_1|X_2,\boldsymbol{S},Q)-H(Z_1)\\ &\leq H(S_{11}X_1+Z_1|\boldsymbol{S})-\log(\pi e)\\ &= \int_{\boldsymbol{S}} \mathbb{P}_{\boldsymbol{S}}(\boldsymbol{s})H(s_{11}X_1+Z_1|\boldsymbol{s})-\log(\pi e)\\ &\leq \int_{\boldsymbol{S}} \mathbb{P}_{\boldsymbol{S}}(\boldsymbol{s})\log\big(\pi e(|s_{11}|^2\mathbb{E}[|X_1|^2|E_1=e_1]+1)\big)-\log(\pi e)\\ &= \mathbb{E}\big[\psi\big(|S_{11}|^2\varphi_1(E_1)\big)\big] \end{aligned} \tag{21}$$

Next, we evaluate the constraints (19, b) and (19, c). One can write:

$$R_1 \leq I(U,X_1;Y_1|\boldsymbol{S},Q) = \int_{|S_{12}|<|S_{22}|} \mathbb{P}_{\boldsymbol{S}}(\boldsymbol{s})I(U,X_1;Y_1|\boldsymbol{s},Q)+\int_{|S_{12}|\geq|S_{22}|} \mathbb{P}_{\boldsymbol{S}}(\boldsymbol{s})I(U,X_1;Y_1|\boldsymbol{s},Q) \tag{22}$$

$$\begin{aligned} &R_1+R_2 \leq I(X_2;Y_2|U,X_1,\boldsymbol{S},Q)+I(U,X_1;Y_1|\boldsymbol{S},Q)\\ &\quad= \int_{|S_{12}|<|S_{22}|} \mathbb{P}_{\boldsymbol{S}}(\boldsymbol{s})\big(I(X_2;Y_2|U,X_1,\boldsymbol{s},Q)+I(U,X_1;Y_1|\boldsymbol{s},Q)\big)+\int_{|S_{12}|\geq|S_{22}|} \mathbb{P}_{\boldsymbol{S}}(\boldsymbol{s})\big(I(X_2;Y_2|U,X_1,\boldsymbol{s},Q)+I(U,X_1;Y_1|\boldsymbol{s},Q)\big) \end{aligned} \tag{23}$$

Consider the first integrals in (22) and (23). Let $\boldsymbol{s} \in \{|S_{12}|<|S_{22}|\}$. We have:

$$\begin{aligned} I(U,X_1;Y_1|\boldsymbol{s},Q) &= H(s_{11}X_1+s_{12}X_2+Z_1|\boldsymbol{s},Q)-H(s_{11}X_1+s_{12}X_2+Z_1|U,X_1,\boldsymbol{s},Q)\\ &= H(s_{11}X_1+s_{12}X_2+Z_1|\boldsymbol{s},Q)-H(s_{12}X_2+Z_1|U,X_1,\boldsymbol{s},Q)\\ &= \mathbb{E}_Q[H(s_{11}X_1+s_{12}X_2+Z_1|\boldsymbol{s},Q)]-H(s_{12}X_2+Z_1|U,X_1,\boldsymbol{s},Q)\\ &\leq \mathbb{E}_Q\Big[\log\Big(\pi e\big(|s_{11}|^2\mathbb{E}\big[|X_1|^2\big|E_1=e_1,Q\big]+|s_{12}|^2\mathbb{E}[|X_2|^2|E_2=e_2,Q]+1\big)\Big)\Big]-H(s_{12}X_2+Z_1|U,X_1,\boldsymbol{s},Q)\\ &\overset{(a)}{\leq} \log\Big(\pi e\big(|s_{11}|^2\mathbb{E}_Q\big[\mathbb{E}[|X_1|^2|E_1=e_1,Q]\big]+|s_{12}|^2\mathbb{E}_Q\big[\mathbb{E}[|X_2|^2|E_2=e_2,Q]\big]+1\big)\Big)-H(s_{12}X_2+Z_1|U,X_1,\boldsymbol{s},Q)\\ &= \log\big(\pi e(|s_{11}|^2\mathbb{E}[|X_1|^2|E_1=e_1]+|s_{12}|^2\mathbb{E}[|X_2|^2|E_2=e_2]+1)\big)-H(s_{12}X_2+Z_1|U,X_1,\boldsymbol{s},Q)\\ &= \log\big(\pi e(|s_{11}|^2\varphi_1(e_1)+|s_{12}|^2\varphi_2(e_2)+1)\big)-H(s_{12}X_2+Z_1|U,X_1,\boldsymbol{s},Q) \end{aligned} \tag{24}$$

where (a) is due to Jensen's inequality. Also,

$$I(X_2;Y_2|U,X_1,\boldsymbol{s},Q)+I(U,X_1;Y_1|\boldsymbol{s},Q)$$

$$= H(s_{21}X_1 + s_{22}X_2 + Z_2|U,X_1,\boldsymbol{s},Q) - H(Z_2) + I(U,X_1;Y_1|\boldsymbol{s},Q)$$
$$\leq H(s_{22}X_2 + Z_2|U,X_1,\boldsymbol{s},Q) - \log(\pi e) + \log\big(\pi e(|s_{11}|^2\varphi_1(e_1) + |s_{12}|^2\varphi_2(e_2) + 1)\big) - H(s_{12}X_2 + Z_1|U,X_1,\boldsymbol{s},Q) \quad (25)$$

Now consider the term $H(s_{22}X_2 + Z_2|U,X_1,\boldsymbol{s},Q)$ in (25). This term can be bounded as:

$$\log(\pi e) = H(Z_2) \leq H(s_{22}X_2 + Z_2|U,X_1,\boldsymbol{s},Q) \leq H(s_{22}X_2 + Z_2|\boldsymbol{s}) \leq \log\big(\pi e(|s_{22}|^2\varphi_2(e_2) + 1)\big) \quad (26)$$

The two sides of the inequality (26) imply that with respect to each $\boldsymbol{s}$ there exist $0 \leq \alpha(\boldsymbol{s}) \leq 1$ such that:

$$H(s_{22}X_2 + Z_2|U,X_1,\boldsymbol{s},Q) = \log\big(\pi e(|s_{22}|^2\alpha(\boldsymbol{s})\varphi_2(e_2) + 1)\big) \quad (27)$$

Now we can bound the term $H(s_{12}X_2 + Z_1|U,X_1,\boldsymbol{s},Q)$ in (24) and (25) using the EPI. Let $\tilde{Z}_1$ be a virtual noise independent of $Z_1$ and $Z_2$ with zero mean and unit variance. We have:

$$\begin{aligned} H(s_{12}X_2 + Z_1|U,X_1,\boldsymbol{s},Q) &= H\left(\frac{s_{12}}{s_{22}}(s_{22}X_2 + Z_2) + \sqrt{1-\left|\frac{s_{12}}{s_{22}}\right|^2}\tilde{Z}_1\middle|U,X_1,\boldsymbol{s},Q\right) \\ &\overset{(a)}{\geq} \log\left(2^{H\left(\frac{s_{12}}{s_{22}}(s_{22}X_2+Z_2)\middle|U,X_1,\boldsymbol{s},Q\right)} + 2^{H\left(\sqrt{1-\left|\frac{s_{12}}{s_{22}}\right|^2}\tilde{Z}_1\middle|U,X_1,\boldsymbol{s},Q\right)}\right) \\ &\overset{(b)}{=} \log\left(\pi e\left|\frac{s_{12}}{s_{22}}\right|^2(|s_{22}|^2\alpha(\boldsymbol{s})\varphi_2(e_2) + 1) + \pi e\left(1-\left|\frac{s_{12}}{s_{22}}\right|^2\right)\right) \\ &= \log\big(\pi e(|s_{12}|^2\alpha(\boldsymbol{s})\varphi_2(e_2) + 1)\big) \end{aligned} \quad (28)$$

where inequality (a) is due to the EPI and equality (b) is derived from (27). Now from (24), (25), (27), and (28), we have:

$$\begin{aligned} \int_{|S_{12}|<|S_{22}|} \mathbb{P}_{\boldsymbol{S}}(\boldsymbol{s}) I(U,X_1;Y_1|\boldsymbol{s},Q) &\leq \int_{|S_{12}|<|S_{22}|} \mathbb{P}_{\boldsymbol{S}}(\boldsymbol{s})\Big(\log\big(\pi e(|s_{11}|^2\varphi_1(e_1) + |s_{12}|^2\varphi_2(e_2) + 1)\big) - \log\big(\pi e(|s_{12}|^2\alpha(\boldsymbol{s})\varphi_2(e_2) + 1)\big)\Big) \\ &= \mathbb{E}\left[\psi\left(\frac{|S_{11}|^2\varphi_1(E_1) + |S_{12}|^2\big(1-\alpha(\boldsymbol{S})\big)\varphi_2(E_2)}{|S_{12}|^2\alpha(\boldsymbol{S})\varphi_2(E_2) + 1}\right)\mathbb{1}(|S_{12}| < |S_{22}|)\right] \end{aligned} \quad (29)$$

$$\begin{aligned} &\int_{|S_{12}|<|S_{22}|} \mathbb{P}_{\boldsymbol{S}}(\boldsymbol{s})\big(I(X_2;Y_2|U,X_1,\boldsymbol{s},Q) + I(U,X_1;Y_1|\boldsymbol{s},Q)\big) \\ &\leq \int_{|S_{12}|<|S_{22}|} \mathbb{P}_{\boldsymbol{S}}(\boldsymbol{s})\begin{pmatrix} \log\big(\pi e(|s_{22}|^2\alpha(\boldsymbol{s})\varphi_2(e_2) + 1)\big) - \log(\pi e) \\ + \log\big(\pi e(|s_{11}|^2\varphi_1(e_1) + |s_{12}|^2\varphi_2(e_2) + 1)\big) - \log\big(\pi e(|s_{12}|^2\alpha(\boldsymbol{s})\varphi_2(e_2) + 1)\big) \end{pmatrix} \\ &= \mathbb{E}\big[\psi\big(|S_{22}|^2\alpha(\boldsymbol{S})\varphi_2(E_2)\big)\mathbb{1}(|S_{12}| < |S_{22}|)\big] + \mathbb{E}\left[\psi\left(\frac{|S_{11}|^2\varphi_1(E_1) + |S_{12}|^2\big(1-\alpha(\boldsymbol{S})\big)\varphi_2(E_2)}{|S_{12}|^2\alpha(\boldsymbol{S})\varphi_2(E_2) + 1}\right)\mathbb{1}(|S_{12}| < |S_{22}|)\right] \end{aligned} \quad (30)$$

Lastly, consider the second integrals in (22) and (23). One can write:

$$\begin{aligned} \int_{|S_{12}|\geq|S_{22}|} \mathbb{P}_{\boldsymbol{S}}(\boldsymbol{s}) I(U,X_1;Y_1|\boldsymbol{s},Q) &\leq \int_{|S_{12}|\geq|S_{22}|} \mathbb{P}_{\boldsymbol{S}}(\boldsymbol{s}) I(X_1,X_2;Y_1|\boldsymbol{s},Q) \\ &\overset{(a)}{\leq} \mathbb{E}\big[\psi\big(|S_{11}|^2\varphi_1(E_1) + |S_{12}|^2\varphi_2(E_2)\big)\mathbb{1}(|S_{12}| \geq |S_{22}|)\big] \end{aligned} \quad (31)$$

where inequality (a) is due to the "Gaussian maximizes the entropy" principle. Also,

$$\int_{|S_{12}|\geq|S_{22}|}\mathbb{P}_{\boldsymbol{S}}(\boldsymbol{s})\big(I(X_2;Y_2|U,X_1,\boldsymbol{s},Q)+I(U,X_1;Y_1|\boldsymbol{s},Q)\big)$$

$$\stackrel{(a)}{\leq}\int_{|S_{12}|\geq|S_{22}|}\mathbb{P}_{\boldsymbol{S}}(\boldsymbol{s})\big(I(X_2;Y_1|U,X_1,\boldsymbol{s},Q)+I(U,X_1;Y_1|\boldsymbol{s},Q)\big)$$

$$=\int_{|S_{12}|\geq|S_{22}|}\mathbb{P}_{\boldsymbol{S}}(\boldsymbol{s})I(X_1,X_2;Y_1|\boldsymbol{s},Q)\leq\mathbb{E}\big[\psi\big(|S_{11}|^2\varphi_1(E_1)+|S_{12}|^2\varphi_2(E_2)\big)\mathbb{1}(|S_{12}|\geq|S_{22}|)\big] \tag{32}$$

where inequality (a) holds because when $\boldsymbol{s}\in\{|S_{12}|\geq|S_{22}|\}$, the receiver $Y_1$ observes strong interference; hence, $I(X_2;Y_2|U,X_1,\boldsymbol{s},Q)\leq I(X_2;Y_1|U,X_1,\boldsymbol{s},Q)$. By substituting (29)-(32) in (22) and (23), we obtain the desired constraints as given in (18).

Next, consider the constraint (19, d). This constraint is evaluated rather similar to (19, c). We can write:

$$R_1+2R_2\leq I(X_2;Y_2|U,X_1,\boldsymbol{S},Q)+I(U,X_1;Y_1|G_1,\boldsymbol{S},Q)+I(X_1,X_2;Y_2|\boldsymbol{S},Q)$$

$$=I(X_2;Y_2|U,X_1,\boldsymbol{S},Q)-H(Y_1|U,X_1,\boldsymbol{S},Q)+H(Y_1|G_1,\boldsymbol{S},Q)+I(X_1,X_2;Y_2|\boldsymbol{S},Q)$$

$$=\int_{|S_{12}|<|S_{22}|}\mathbb{P}_{\boldsymbol{S}}(\boldsymbol{s})\big(I(X_2;Y_2|U,X_1,\boldsymbol{s},Q)-H(Y_1|U,X_1,\boldsymbol{s},Q)\big)+\int_{|S_{12}|\geq|S_{22}|}\mathbb{P}_{\boldsymbol{S}}(\boldsymbol{s})\big(I(X_2;Y_2|U,X_1,\boldsymbol{s},Q)-H(Y_1|U,X_1,\boldsymbol{S},Q)\big)$$

$$+H(Y_1|G_1,\boldsymbol{S},Q)+I(X_1,X_2;Y_2|\boldsymbol{S},Q) \tag{33}$$

Now, for the first expression of (33) we have:

$$\int_{|S_{12}|<|S_{22}|}\mathbb{P}_{\boldsymbol{S}}(\boldsymbol{s})\big(I(X_2;Y_2|U,X_1,\boldsymbol{s},Q)-H(Y_1|U,X_1,\boldsymbol{s},Q)\big)$$

$$=\int_{|S_{12}|<|S_{22}|}\mathbb{P}_{\boldsymbol{S}}(\boldsymbol{s})\big(H(Y_2|U,X_1,\boldsymbol{s},Q)-\log(\pi e)-H(Y_1|U,X_1,\boldsymbol{s},Q)\big)$$

$$\stackrel{(a)}{\leq}\int_{|S_{12}|<|S_{22}|}\mathbb{P}_{\boldsymbol{S}}(\boldsymbol{s})\big(\log\big(\pi e(|s_{22}|^2\alpha(\boldsymbol{s})\varphi_2(e_2)+1)\big)-\log(\pi e)-\log\big(\pi e(|s_{12}|^2\alpha(\boldsymbol{s})\varphi_2(e_2)+1)\big)\big)$$

$$=\mathbb{E}\big[\psi\big(|S_{22}|^2\alpha(\boldsymbol{S})\varphi_2(E_2)\big)\mathbb{1}(|S_{12}|<|S_{22}|)\big]-\mathbb{E}\big[\psi\big(|S_{12}|^2\alpha(\boldsymbol{S})\varphi_2(E_2)\big)\mathbb{1}(|S_{12}|<|S_{22}|)\big]-\log(\pi e)\int_{|S_{12}|<|S_{22}|}\mathbb{P}_{\boldsymbol{S}}(\boldsymbol{s}) \tag{34}$$

where inequality (a) is due to (27) and (28). For the second expression of (33), one can write:

$$\int_{|S_{12}|\geq|S_{22}|}\mathbb{P}_{\boldsymbol{S}}(\boldsymbol{s})\big(I(X_2;Y_2|U,X_1,\boldsymbol{s},Q)-H(Y_1|U,X_1,\boldsymbol{S},Q)\big)$$

$$\stackrel{(a)}{\leq}\int_{|S_{12}|\geq|S_{22}|}\mathbb{P}_{\boldsymbol{S}}(\boldsymbol{s})\big(I(X_2;Y_1|U,X_1,\boldsymbol{s},Q)-H(Y_1|U,X_1,\boldsymbol{S},Q)\big)$$

$$=\int_{|S_{12}|\geq|S_{22}|}\mathbb{P}_{\boldsymbol{S}}(\boldsymbol{s})\big(-H(Y_1|U,X_1,X_2,\boldsymbol{s},Q)\big)=\int_{|S_{12}|\geq|S_{22}|}\mathbb{P}_{\boldsymbol{S}}(\boldsymbol{s})\big(-H(Z_1)\big)=-\log(\pi e)\int_{|S_{12}|\geq|S_{22}|}\mathbb{P}_{\boldsymbol{S}}(\boldsymbol{s}) \tag{35}$$

where inequality (a) holds because when $\boldsymbol{s}\in\{|S_{12}|\geq|S_{22}|\}$, the receiver $Y_1$ observes strong interference; hence, $I(X_2;Y_2|U,X_1,\boldsymbol{s},Q)\leq I(X_2;Y_1|U,X_1,\boldsymbol{s},Q)$. Then, consider the third term of (33). We have:

$$H(Y_1|G_1,\boldsymbol{S},Q)=\mathbb{E}_Q\mathbb{E}_{\boldsymbol{S}}[H(S_{11}X_1+S_{12}X_2+Z_1|S_{21}X_1+Z_2,\boldsymbol{S},Q)]$$

$$
\begin{aligned}
&\overset{(a)}{\leq} \mathbb{E}_Q \mathbb{E}_{\boldsymbol{S}}\left[\log\left(\pi e\left(1+|S_{12}|^2\mathbb{E}[|X_2|^2|E_2,Q]+\frac{|S_{11}|^2\mathbb{E}[|X_1|^2|E_1,Q]}{|S_{21}|^2\mathbb{E}[|X_1|^2|E_1,Q]+1}\right)\right)\right]\\
&\overset{(b)}{\leq} \mathbb{E}_{\boldsymbol{S}}\left[\log\left(\pi e\left(1+|S_{12}|^2\mathbb{E}_Q\big[\mathbb{E}[|X_2|^2|E_2,Q]\big]+\frac{|S_{11}|^2\mathbb{E}_Q\big[\mathbb{E}[|X_1|^2|E_1,Q]\big]}{|S_{21}|^2\mathbb{E}_Q\big[\mathbb{E}[|X_1|^2|E_1,Q]\big]+1}\right)\right)\right]\\
&= \mathbb{E}\left[\psi\left(|S_{12}|^2\varphi_2(E_2)+\frac{|S_{11}|^2\varphi_1(E_1)}{|S_{21}|^2\varphi_1(E_1)+1}\right)\right]+\log(\pi e)
\end{aligned}
\tag{36}
$$

where (a) is due to the "Gaussian maximizes the entropy" principle, and (b) is due to Jensen's inequality. Also, the fourth term of (33) is directly bounded as:

$$
I(X_1,X_2;Y_2|\boldsymbol{S},Q) \leq \mathbb{E}\big[\psi\big(|S_{21}|^2\varphi_1(E_1)+|S_{22}|^2\varphi_2(E_2)\big)\big]
\tag{37}
$$

Thus, by substituting (34-37) in (33), we obtain the desired constraint as given in (18). Finally, consider the constraint (19, e). This constraint is also evaluated as follows:

$$
\begin{aligned}
R_1+R_2 &\leq I(X_1,X_2;Y_1|G_1,\boldsymbol{S},Q)+I(X_1,X_2;Y_2|G_2,\boldsymbol{S},Q)\\
&= H(Y_1|G_1,\boldsymbol{S},Q)-H(Z_1)+H(Y_2|G_2,\boldsymbol{S},Q)-H(Z_2)\\
&\overset{(a)}{\leq} \mathbb{E}\left[\psi\left(|S_{12}|^2\varphi_2(E_2)+\frac{|S_{11}|^2\varphi_1(E_1)}{|S_{21}|^2\varphi_1(E_1)+1}\right)\right]+\mathbb{E}\left[\psi\left(|S_{21}|^2\varphi_1(E_1)+\frac{|S_{22}|^2\varphi_2(E_2)}{|S_{12}|^2\varphi_2(E_2)+1}\right)\right]
\end{aligned}
\tag{38}
$$

where (a) is derived by following the same lines as (36). The proof is thus complete. ■

***Remarks 3:***

1. The outer bound $\mathfrak{R}_o^{GFIC}$ given in (18) is uniformly applicable for all channels with arbitrary fading statistics and arbitrary amount of state information at the transmitters.
2. For the case where the state process is i.i.d., in the characterization of the outer bound $\mathfrak{R}_o^{GFIC}$ in (18) one can restrict the functions $\alpha(.)$ and $\beta(.)$ to depend only on $(S_{22},E_2)$ and $(S_{11},E_1)$, respectively, and not on the whole of the state $\boldsymbol{S}=\begin{bmatrix}S_{11} & S_{12}\\ S_{21} & S_{22}\end{bmatrix}$. Specially, if there is no side information at the transmitters, i.e., $E_1\equiv E_2\equiv\emptyset$, then $\alpha(.)$ and $\beta(.)$ are decreasing functions[3] of $S_{22}$ and $S_{11}$, respectively. This is due to the derivation (27).

Before analyzing the outer bound (18) for the Gaussian fading channel (1), let us discuss its characteristics for the non-fading case where the channel gains $S_{11},S_{12},S_{21}$ and $S_{22}$ are fixed (deterministic) complex numbers. In this case, without loss of generality, one can assume that $S_{11}\equiv S_{22}\equiv 1$. It can be easily shown that if the channel is in the strong interference regime, i.e., $|S_{21}|\geq 1$ and $|S_{12}|\geq 1$, then the outer bound (18) is optimal. Now, consider the channel with weak interference where $|S_{21}|<1$ and $|S_{12}|<1$. For this scenario, the outer bound $\mathfrak{R}_o^{GFIC}$ in (18) is reduced to the following rate region:

[3] A deterministic function $f(.):\mathbb{C}\rightarrow[0,1]$ is said to be decreasing if $\forall a,b: |a|\geq|b| \Longrightarrow f(a)\leq f(b)$.

$$\bigcup_{\alpha,\beta}\left\{\begin{array}{l}(R_1,R_2)\in\mathbb{R}_+^2:\\ R_1\le\min\left\{\psi(P_1),\psi\left(\frac{P_1+|S_{12}|^2(1-\alpha)P_2}{|S_{12}|^2\alpha P_2+1}\right)\right\}\\ R_2\le\min\left\{\psi(P_2),\psi\left(\frac{|S_{21}|^2(1-\beta)P_1+P_2}{|S_{21}|^2\beta P_1+1}\right)\right\}\\ R_1+R_2\le\psi(\beta P_1)+\psi\left(\frac{|S_{21}|^2(1-\beta)P_1+P_2}{|S_{21}|^2\beta P_1+1}\right)\\ R_1+R_2\le\psi(\alpha P_2)+\psi\left(\frac{P_1+|S_{12}|^2(1-\alpha)P_2}{|S_{12}|^2\alpha P_2+1}\right)\\ R_1+R_2\le\psi\left(|S_{12}|^2P_2+\frac{P_1}{|S_{21}|^2P_1+1}\right)+\psi\left(|S_{21}|^2P_1+\frac{P_2}{|S_{12}|^2P_2+1}\right)\\ 2R_1+R_2\le\psi(\beta P_1)-\psi(|S_{21}|^2\beta P_1)+\psi\left(|S_{21}|^2P_1+\frac{P_2}{|S_{12}|^2P_2+1}\right)+\psi(P_1+|S_{12}|^2P_2)\\ R_1+2R_2\le\psi(\alpha P_2)-\psi(|S_{12}|^2\alpha P_2)+\psi\left(|S_{12}|^2P_2+\frac{P_1}{|S_{21}|^2P_1+1}\right)+\psi(|S_{21}|^2P_1+P_2)\end{array}\right\}$$

(39)

We can analytically show that this outer bound is strictly tighter than both the Kramer's outer bound [32] and the ETW outer bound [7, Th. 3]. Clearly, if we consider only the first four constraints of (39) and relax the others, the resultant region is equivalent to the Kramer's outer bound [32]. In fact, the Kramer's outer bound can be simply re-derived by evaluating the UV-outer bound established in our recent paper for the IC [2, Sec. III.B][4]. Also, if we relax the following two constraints:

$$\begin{cases}R_1\le\psi\left(\frac{P_1+|S_{12}|^2(1-\alpha)P_2}{|S_{12}|^2\alpha P_2+1}\right)\\ R_2\le\psi\left(\frac{|S_{21}|^2(1-\beta)P_1+P_2}{|S_{21}|^2\beta P_1+1}\right)\end{cases}$$

(40)

then, one can easily verify that the remaining constraints are optimized for $\alpha=\beta=1$, and the resultant region is equal to the ETW outer bound [7, Th. 3]. Note that the constraints (40) are actually derived by evaluating the following constrains of $\mathfrak{R}_o^{UV\to GFIC}$ in (6):

$$\begin{cases}R_1\le I(U,X_1;Y_1|Q)\\ R_2\le I(V,X_2;Y_2|Q)\end{cases}$$

(41)

If these constraints are relaxed from the outer bound (6), then the resultant region is optimized (provided that the channel has weak interference) for $U\equiv V\equiv\emptyset$ which results to the ETW outer bound [7, Th. 3]. This indicates that the inclusion of the auxiliaries $U$ and $V$ into the constraints given in (6) on the linear combinations $2R_1+R_2$ and $R_1+2R_2$ makes them tighter than their counterparts established in [7, Th. 3] using the worst case noise lemma [27].

For the Gaussian mixed IC where $|S_{21}|<1$ and $|S_{12}|\ge 1$ (or $|S_{21}|<1$ and $|S_{12}|\ge 1$) again our outer bound is strictly tighter than the ETW bound [7, Th. 4]. In fact, for this case even if we relax the constraints (41) from the characterization (6), the resultant bound is still tighter than that of [7, Th. 4]. This is because for a mixed Gaussian IC with $|S_{21}|<1$ and $|S_{12}|\ge 1$, the constraint of our outer bound (18) on $R_1+2R_2$ is reduced to:

$$R_1+2R_2\le\psi\left(|S_{12}|^2P_2+\frac{P_1}{|S_{21}|^2P_1+1}\right)+\psi(|S_{21}|^2P_1+P_2)$$

(42)

while the counterpart of this constraint in the bound of [7, Th. 4] is given by:

[4] We remark that the UV-outer bound for the IC is indeed obtained inspired by its counterpart for the two-user BC as described in details in our paper [2, Sec. III.B]. An alternative characterization of the Kramer's outer bound is also given in [23, Lemma 11], however, it is complex and not directly adaptable for the fading channel.

$$R_1+2R_2\le\psi\left(|S_{12}|^2P_2+\frac{P_1}{|S_{21}|^2P_1+1}\right)+\psi(|S_{21}|^2P_1+P_2)+\psi\left(\frac{P_2}{|S_{12}|^2P_2+1}\right)\tag{43}$$

Note that (43) differs from (42) by the additional term $\psi\left(\frac{P_2}{|S_{12}|^2P_2+1}\right)$.

Let return to the Gaussian fading channel (1). In general, it is rather difficult to explicitly determine the optimal functions $\alpha(.)$ and $\beta(.)$ that achieves the boundary points of the outer bound $\mathfrak{R}_o^{GFIC}$ in (18). However, for the purposes of this paper we need not to do so. Clearly, if we relax the following constraints from (18):

$$\begin{cases} R_1\le I(U,X_1;Y_1|\boldsymbol{S},Q)\le\mathbb{E}\left[\psi\left(\frac{|S_{11}|^2\varphi_1(E_1)+|S_{12}|^2\big(1-\alpha(\boldsymbol{S})\big)\varphi_2(E_2)}{|S_{12}|^2\alpha(\boldsymbol{S})\varphi_2(E_2)+1}\right)\mathbb{1}(|S_{12}|<|S_{22}|)\right] \\ \qquad +\mathbb{E}\left[\psi\big(|S_{11}|^2\varphi_1(E_1)+|S_{12}|^2\varphi_2(E_2)\big)\mathbb{1}(|S_{12}|\ge|S_{22}|)\right] \\ , \\ R_2\le I(V,X_2;Y_2|\boldsymbol{S},Q)\le\mathbb{E}\left[\psi\left(\frac{|S_{21}|^2\big(1-\beta(\boldsymbol{S})\big)\varphi_1(E_1)+|S_{22}|^2\varphi_2(E_2)}{|S_{21}|^2\beta(\boldsymbol{S})\varphi_1(E_1)+1}\right)\mathbb{1}(|S_{21}|<|S_{11}|)\right] \\ \qquad +\mathbb{E}\left[\psi\big(|S_{21}|^2\varphi_1(E_1)+|S_{22}|^2\varphi_2(E_2)\big)\mathbb{1}(|S_{21}|\ge|S_{11}|)\right] \end{cases}\tag{44}$$

then, the remaining constraints are optimized for $\alpha(.)\equiv\beta(.)\equiv1$. The resultant region, which we denote by $\mathfrak{R}_{o:(2)}^{GFIC}$, is given as follows:

$$\mathfrak{R}_{o:(2)}^{GFIC}\triangleq\bigcup_{\substack{\varphi_1(.),\varphi_2(.)\\ \mathbb{E}[\varphi_i(E_i)]\le P_i,i=1,2}}\left\{\begin{array}{l} (R_1,R_2)\in\mathbb{R}_+^2: \\ R_1\le\mathbb{E}\left[\psi\big(|S_{11}|^2\varphi_1(E_1)\big)\right] \\ R_2\le\mathbb{E}\left[\psi\big(|S_{22}|^2\varphi_2(E_2)\big)\right] \\ R_1+R_2\le\mathbb{E}\left[\left(\psi\big(|S_{11}|^2\varphi_1(E_1)\big)+\psi\left(\frac{|S_{22}|^2\varphi_2(E_2)}{|S_{21}|^2\varphi_1(E_1)+1}\right)\right)\mathbb{1}(|S_{21}|<|S_{11}|)\right] \\ \qquad +\mathbb{E}\left[\psi\big(|S_{21}|^2\varphi_1(E_1)+|S_{22}|^2\varphi_2(E_2)\big)\mathbb{1}(|S_{21}|\ge|S_{11}|)\right] \\ , \\ R_1+R_2\le\mathbb{E}\left[\left(\psi\big(|S_{22}|^2\varphi_2(E_2)\big)+\psi\left(\frac{|S_{11}|^2\varphi_1(E_1)}{|S_{12}|^2\varphi_2(E_2)+1}\right)\right)\mathbb{1}(|S_{12}|<|S_{22}|)\right] \\ \qquad +\mathbb{E}\left[\psi\big(|S_{11}|^2\varphi_1(E_1)+|S_{12}|^2\varphi_2(E_2)\big)\mathbb{1}(|S_{12}|\ge|S_{22}|)\right] \\ , \\ R_1+R_2\le\mathbb{E}\left[\psi\left(|S_{12}|^2\varphi_2(E_2)+\frac{|S_{11}|^2\varphi_1(E_1)}{|S_{21}|^2\varphi_1(E_1)+1}\right)\right]+\mathbb{E}\left[\psi\left(|S_{21}|^2\varphi_1(E_1)+\frac{|S_{22}|^2\varphi_2(E_2)}{|S_{12}|^2\varphi_2(E_2)+1}\right)\right] \\ , \\ 2R_1+R_2\le\mathbb{E}\left[\Big(\psi\big(|S_{11}|^2\varphi_1(E_1)\big)-\psi\big(|S_{21}|^2\varphi_1(E_1)\big)\Big)\mathbb{1}(|S_{21}|<|S_{11}|)\right] \\ \qquad +\mathbb{E}\left[\psi\left(|S_{21}|^2\varphi_1(E_1)+\frac{|S_{22}|^2\varphi_2(E_2)}{|S_{12}|^2\varphi_2(E_2)+1}\right)\right]+\mathbb{E}\left[\psi\big(|S_{11}|^2\varphi_1(E_1)+|S_{12}|^2\varphi_2(E_2)\big)\right] \\ R_1+2R_2\le\mathbb{E}\left[\Big(\psi\big(|S_{22}|^2\varphi_2(E_2)\big)-\psi\big(|S_{12}|^2\varphi_2(E_2)\big)\Big)\mathbb{1}(|S_{12}|<|S_{22}|)\right] \\ \qquad +\mathbb{E}\left[\psi\left(|S_{12}|^2\varphi_2(E_2)+\frac{|S_{11}|^2\varphi_1(E_1)}{|S_{21}|^2\varphi_1(E_1)+1}\right)\right]+\mathbb{E}\left[\psi\big(|S_{21}|^2\varphi_1(E_1)+|S_{22}|^2\varphi_2(E_2)\big)\right] \end{array}\right\}\tag{45}$$

This outer bound is in general strictly weaker than $\mathfrak{R}_o^{GFIC}$ in (18), however, it is sufficient to derive our desired results in the rest of the paper. Here, we remark that an outer bound was also reported in [1] for the general Gaussian fading IC with perfect CSIT. Note that the outer bound of [1] does not include any constraint on $2R_1+R_2$ and $R_1+2R_2$. Even without considering these constraints, by a simple comparison (equation by equation) one can analytically show that our outer bound in (45) is strictly tighter than that of [1].

We next prove some important results for the channel using the outer bound $\mathfrak{R}_{o:(2)}^{GFIC}$ in (45). First we show that for the case where each transmitter has access to the interference to noise ratio perceived at its non-corresponding receiver, this outer bound is to within one bit of the capacity region. This result is given in the following theorem.

***Theorem 2)*** *Consider the two-user Gaussian fading IC (1) with partial side information $E_1$ and $E_2$ at the transmitters $X_1$ and $X_2$, respectively. Assume that $E_1 \equiv (|S_{21}|, E_1^*)$ and $E_2 \equiv (|S_{12}|, E_2^*)$ where $E_1^*$ and $E_2^*$ are given by arbitrary deterministic functions of the channel state $\boldsymbol{S}$. If $(R_1, R_2)$ belongs to the outer bound $\mathfrak{R}_{o:(2)}^{GFIC}$ in (45), then $(R_1 - 1, R_2 - 1)$ is achievable.*

*Proof of Theorem 2)* Fix two deterministic functions $\varphi_1(.)$ and $\varphi_2(.)$ with $\mathbb{E}[\varphi_i(E_i)] \le P_i, i = 1,2$. Let $\mathfrak{R}_{o:(2)}^{GFIC}(\varphi_1(E_1), \varphi_2(E_2))$ denotes a subset of $\mathfrak{R}_{o:(2)}^{GFIC}$ that is given by equation (45) for the fixed functions $\varphi_1(.)$ and $\varphi_2(.)$. Also, let $\mathfrak{R}_i^{GFIC}(\varphi_1(E_1), \varphi_2(E_2))$ be a subset of $\mathfrak{R}_i^{GFIC}$ in (2) that is determined by the power allocation policies $\varphi_1(.)$ and $\varphi_2(.)$ and the following assignments:

$$\alpha(E_1) = \alpha(|S_{21}|, E_1^*) \triangleq \min\left\{1, \frac{1}{|S_{21}|^2\varphi_1(E_1)}\right\}, \qquad \beta(E_2) = \beta(|S_{12}|, E_2^*) \triangleq \min\left\{1, \frac{1}{|S_{12}|^2\varphi_2(E_2)}\right\} \tag{46}$$

We show that if $(R_1, R_2)$ belongs to $\mathfrak{R}_{o:(2)}^{GFIC}(\varphi_1(E_1), \varphi_2(E_2))$, then $(R_1 - 1, R_2 - 1)$ belongs to $\mathfrak{R}_i^{GFIC}(\varphi_1(E_1), \varphi_2(E_2))$. This can be proved by an equation-by-equation comparison of the latter regions. We do this only for the corresponding constraints on $R_1 + 2R_2$. The other constraints can be compared similarly. For the region $\mathfrak{R}_{o:(2)}^{GFIC}(\varphi_1(E_1), \varphi_2(E_2))$, the linear combination $R_1 + 2R_2$ is bounded above as:

$$R_1 + 2R_2 \le A_o + B_o + C_o \tag{47}$$

where:

$$\begin{cases} A_o = \mathbb{E}\left[\left(\psi(|S_{22}|^2\varphi_2(E_2)) - \psi(|S_{12}|^2\varphi_2(E_2))\right)\mathbb{1}(|S_{12}| < |S_{22}|)\right] \\ B_o = \mathbb{E}\left[\psi\left(|S_{12}|^2\varphi_2(E_2) + \dfrac{|S_{11}|^2\varphi_1(E_1)}{|S_{21}|^2\varphi_1(E_1) + 1}\right)\right] \\ C_o = \mathbb{E}\left[\psi(|S_{21}|^2\varphi_1(E_1) + |S_{22}|^2\varphi_2(E_2))\right] \end{cases} \tag{48}$$

On the one hand, for the region $\mathfrak{R}_i^{GFIC}(\varphi_1(E_1), \varphi_2(E_2))$ it is bounded above as:

$$R_1 + 2R_2 \le A_i + B_i + C_i \tag{49}$$

where:

$$\begin{cases} A_i = \mathbb{E}\left[\psi\left(\dfrac{|S_{22}|^2\varphi_2(E_2)\beta(E_2)}{|S_{21}|^2\varphi_1(E_1)\alpha(E_1) + 1}\right)\right] \\ B_i = \mathbb{E}\left[\psi\left(\dfrac{|S_{11}|^2\varphi_1(E_1)\alpha(E_1) + |S_{12}|^2\varphi_2(E_2)(1 - \beta(E_2))}{|S_{12}|^2\varphi_2(E_2)\beta(E_2) + 1}\right)\right] \\ C_i = \mathbb{E}\left[\psi\left(\dfrac{|S_{21}|^2\varphi_1(E_1)(1 - \alpha(E_1)) + |S_{22}|^2\varphi_2(E_2)}{|S_{21}|^2\varphi_1(E_1)\alpha(E_1) + 1}\right)\right] \end{cases} \tag{50}$$

Note that $\alpha(E_1)$ and $\beta(E_2)$ are given by (46). Now, it is sufficient to show:

$$A_o \le A_i + 1, \qquad B_o \le B_i + 1, \qquad C_o \le C_i + 1 \tag{51}$$

Consider the first inequality of (51). We have:

$$A_o = \mathbb{E}\left[\left(\log\left(\frac{1 + |S_{22}|^2\varphi_2(E_2)}{1 + |S_{12}|^2\varphi_2(E_2)}\right)\right)\mathbb{1}(|S_{12}| < |S_{22}|)\right]$$

$$\le \mathbb{E}\left[\left(\log\left(\beta(E_2)(1 + |S_{22}|^2\varphi_2(E_2))\right)\right)\mathbb{1}(|S_{12}| < |S_{22}|)\right]$$

$$\leq \mathbb{E}\left[\log\left(\beta(E_2) + |S_{22}|^2\varphi_2(E_2)\beta(E_2)\right)\right]$$
$$\leq \mathbb{E}\left[\log\left(2 + |S_{22}|^2\varphi_2(E_2)\beta(E_2)\right)\right]$$
$$= \mathbb{E}\left[\log\left(1 + \frac{|S_{22}|^2\varphi_2(E_2)\beta(E_2)}{2}\right)\right] + 1$$
$$\leq \mathbb{E}\left[\log\left(1 + \frac{|S_{22}|^2\varphi_2(E_2)\beta(E_2)}{|S_{21}|^2\varphi_1(E_1)\alpha(E_1) + 1}\right)\right] + 1 = A_i + 1 \tag{52}$$

The two other inequalities of (51) can also be proved by similar algebraic computations. The proof is thus complete. ■

***Remarks 4:***

1. The result of Theorem 2 holds for arbitrary stationary and ergodic fading statistics.
2. From the viewpoint of CSIT quality, Theorem 2 holds for a broad range of possible scenarios because the only requirement is that the transmitters $X_1$ and $X_2$ have access to $|S_{21}|$ and $|S_{12}|$, respectively. It is indeed interesting that to achieve this "to within one bit" capacity result, we need not to impose the restrictive assumption of prefect CSIT (although the result clearly holds for this case, as well).
3. The consequence of Theorem 2 indeed provides a clear understating of the two-user wireless ergodic fading IC for the case where each transmitter has access to the interference to noise ratio perceived at its non-corresponding receiver.

Then, we present some special cases for which the outer bound $\mathfrak{R}_{o:(2)}^{GFIC}$ in (45) is optimal. Specifically, we consider the class of *uniformly strong* as well as *uniformly mixed* Gaussian fading ICs studied in the recent work [11].

***Definition:*** *The two-user Gaussian fading IC (1) is said to have uniformly strong interference provided that:*

$$Pr\{|S_{11}| > |S_{21}|\} = Pr\{|S_{22}| > |S_{12}|\} = 0 \tag{53}$$

*In other words, both receivers at each time instant perceive strong interference with probability 1.*

For the case where both transmitters have access to perfect state information, the capacity region of the channel with uniformly strong interference was established in [11]. It was shown that decoding both messages at both receivers is optimal.

***Definition:*** *The two-user Gaussian fading IC (2) is said to have uniformly mixed interference provided that:*

$$Pr\{|S_{11}| < |S_{21}|\} = Pr\{|S_{22}| > |S_{12}|\} = 0 \tag{54}$$

*In other words, with probability 1, the receiver $Y_1$ perceives strong interference and the receiver $Y_2$ perceives weak interference.*

The authors in [11] also established the sum-rate capacity of the fading IC with uniformly mixed interference for the case where both transmitters have access to perfect state information. The optimal strategy to achieve the sum-rate capacity is that the receiver $Y_1$ fully decodes the interference (it decodes both messages) while the receiver $Y_2$ treats interference as noise (it only decodes its own message). The converse part was proved by bounding the sum-rate capacity of the channel by the sum-capacities of the corresponding strong and weak one-sided channels which result from eliminating one of the two interfering links.

In the following, we extend the capacity results of [11] to the case with arbitrary amount of state information at the transmitters. For this purpose, we exploit our outer bound given by (45).

***Proposition 2)*** *Consider the two-user Gaussian fading IC (1) with partial CSIT.*

*I. The outer bound $\mathfrak{R}_{o:(2)}^{GFIC}$ in (45) is optimal for the channel with uniformly strong interference (53). The capacity region is given by:*

$$\bigcup_{\substack{\varphi_1(.),\varphi_2(.) \\ \mathbb{E}[\varphi_i(E_i)]\leq P_i, i=1,2}} \left\{\begin{array}{l} (R_1,R_2)\in\mathbb{R}_+^2: \\ R_1 \leq \mathbb{E}\big[\psi\big(|S_{11}|^2\varphi_1(E_1)\big)\big] \\ R_2 \leq \mathbb{E}\big[\psi\big(|S_{22}|^2\varphi_2(E_2)\big)\big] \\ R_1+R_2 \leq \mathbb{E}\big[\psi\big(|S_{11}|^2\varphi_1(E_1)+|S_{12}|^2\varphi_2(E_2)\big)\big] \\ R_1+R_2 \leq \mathbb{E}\big[\psi\big(|S_{21}|^2\varphi_1(E_1)+|S_{22}|^2\varphi_2(E_2)\big)\big] \end{array}\right\} \tag{55}$$

II. *The outer bound* $\mathfrak{R}_{o:(2)}^{GFIC}$ *in (45) is sum-rate optimal for the channel with mixed interference (54). The sum-rate capacity is given by:*

$$\max_{\substack{\varphi_1(.),\varphi_2(.) \\ \mathbb{E}[\varphi_i(E_i)]\leq P_i, i=1,2}} \min\left\{\begin{array}{c} \mathbb{E}\left[\left(\psi\big(|S_{11}|^2\varphi_1(E_1)\big)+\psi\left(\dfrac{|S_{22}|^2\varphi_2(E_2)}{|S_{21}|^2\varphi_1(E_1)+1}\right)\right)\right], \\ \mathbb{E}\big[\psi\big(|S_{11}|^2\varphi_1(E_1)+|S_{12}|^2\varphi_2(E_2)\big)\big] \end{array}\right\} \tag{56}$$

*Proof of Proposition 2)*

Part I) The achievability is obtained by viewing the channel as a *compound MAC*, i.e., each receiver decodes both messages. We remark that the capacity region of the compound MAC with partial state information at the transmitters can be obtained as a special case of [19, Th. 1]. Alternatively, one can derive the achievability of (55) for the channel with strong interference (53) by setting $\alpha(.)\equiv\beta(.)\equiv 0$ in the general achievable rate region (2). To prove the converse part, one can easily see that for the channel with strong interference (53), the first four constraints of the outer bound $\mathfrak{R}_{o:(2)}^{GFIC}$ in (45) are reduced to (56).

Part II) The achievability is readily derived by a simple signaling scheme in which the receiver $Y_1$ decodes both transmitted messages and the receiver $Y_2$ only decodes its own message. To prove the converse part, consider the outer bound $\mathfrak{R}_{o:(2)}^{GFIC}$ in (45). For the channel with uniformly mixed interference (54), the third and the fourth constraints of this bound are reduced to (56). The proof is thus complete. ■

We finally provide some numerical examples to compare the derived bounds. Consider the channel with no CSIT, i.e., $E_1\equiv E_2\equiv\emptyset$. Assume that $|S_{11}|, |S_{12}|, |S_{21}|$, and $|S_{22}|$ are independent *Rayleigh-distributed* random variables with the parameters $1, .15, .15$, and $1$, respectively. Also, assume that the transmitters are subject to equal power constraints: $P_1=P_2=P$. Figure 2 compare the inner bound $\mathfrak{R}_i^{GFIC}$ in (2) and the outer bound $\mathfrak{R}_{o:(2)}^{GFIC}$ in (45) for three values of $P$. For each case, the red curve depicts the inner bound and the blue curve depicts the outer bound.

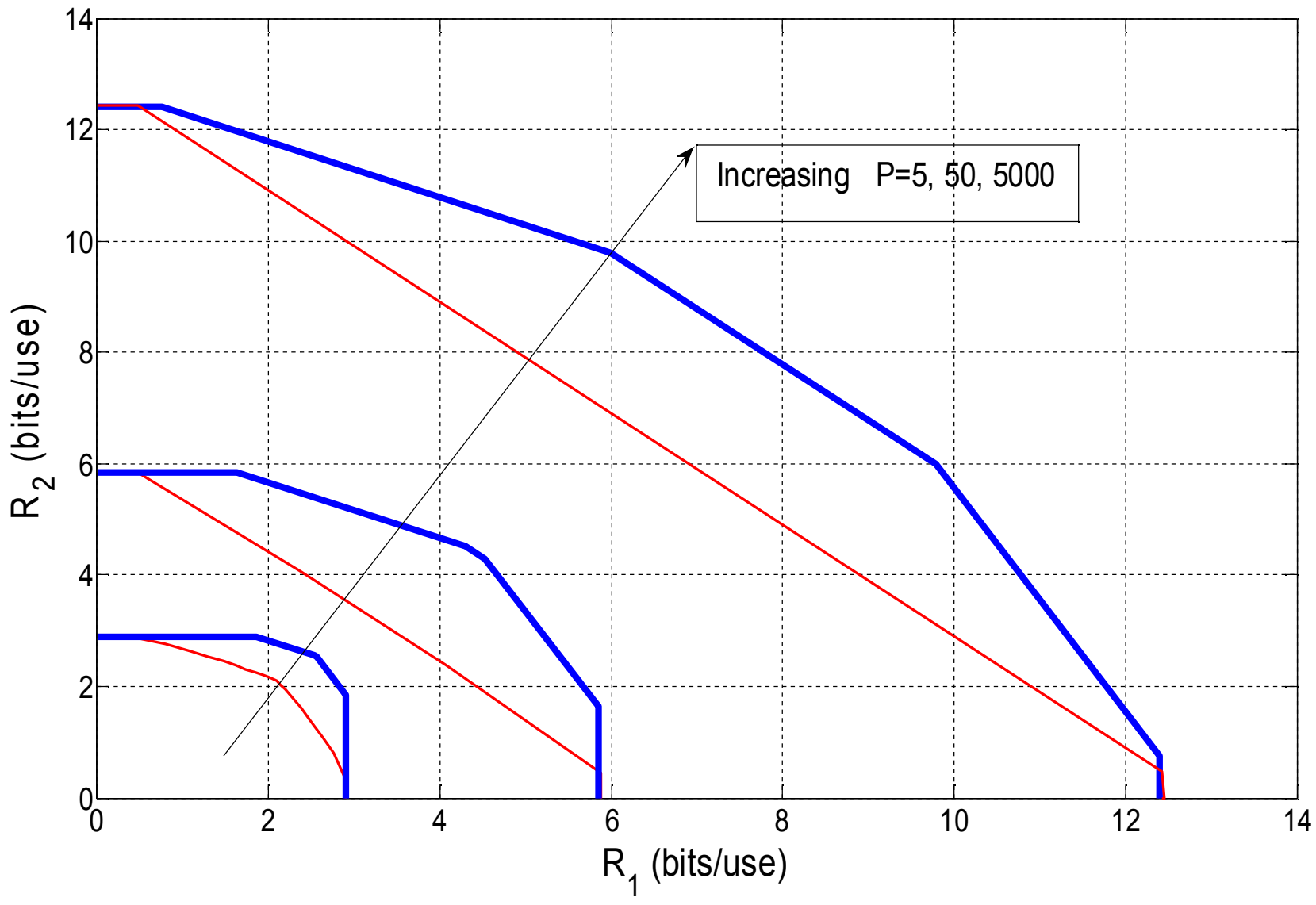

Figure 2. Comparison between the inner bound $\mathfrak{R}_i^{GFIC}$ in (2) and the outer bound $\mathfrak{R}_{o:(2)}^{GFIC}$ in (45) for a Rayleigh fading IC with no CSIT.

## *Discussion on the "to within one bit" capacity result:*

In Theorem 2 we proved that for the wireless ergodic fading channel (1) if each transmitter has access to the interference to noise ratio of its non-corresponding receiver, our outer bound in (45) (or the inner bound (2)) is to within one bit of the capacity region. It is indeed viewed as a natural generalization of the ETW result for the static channel [7] to the wireless ergodic fading case. In the derivation of outer bounds that characterize the capacity region to within one bit, the authors of [7] are inspired by the injective deterministic IC introduced in [33]. In fact, by identifying a connection between the Gaussian channel and the injective deterministic channel and based on the converse proof of [33] for the capacity region of the latter channel, the authors in [7] could recognize appropriate genie signals that should be given to the receivers as side information to obtain desired outer bounds for the Gaussian IC. Later in [34], a class of injective semi-deterministic IC was introduced that includes both the Gaussian channel and the injective deterministic channel as special cases. For this channel, the authors of [34] established an outer bound (inspired by the injective deterministic channel) and showed that it is within a gap specified in terms of certain mutual information functions to the general HK achievable rate region. For the Gaussian channel, this gap is no more than one bit. It should be noted that the approach of [34] is more concise than the original proof of [7], however, they are in spirit similar to each other. A special case of our result in Theorem 2 for the Gaussian fading IC (1) can also be derived by generalizing the approach of [34] for state-dependent channels. For this purpose, we need to develop a class of ***state-dependent injective deterministic IC***. Clearly, consider the two-user state-dependent IC depicted in Fig. 3. In this scenario, $S$ denotes the channel state, $X_1$ and $X_2$ are the inputs, $E_1 = \xi_1(S)$ and $E_2 = \xi_2(S)$ are the side information at the transmitters, and $Y_1$ and $Y_2$ are the channel outputs which are given by the functions:

$$Y_1 = f_1(X_1, G_2, S), \qquad Y_2 = f_2(X_2, G_1, S) \tag{57}$$

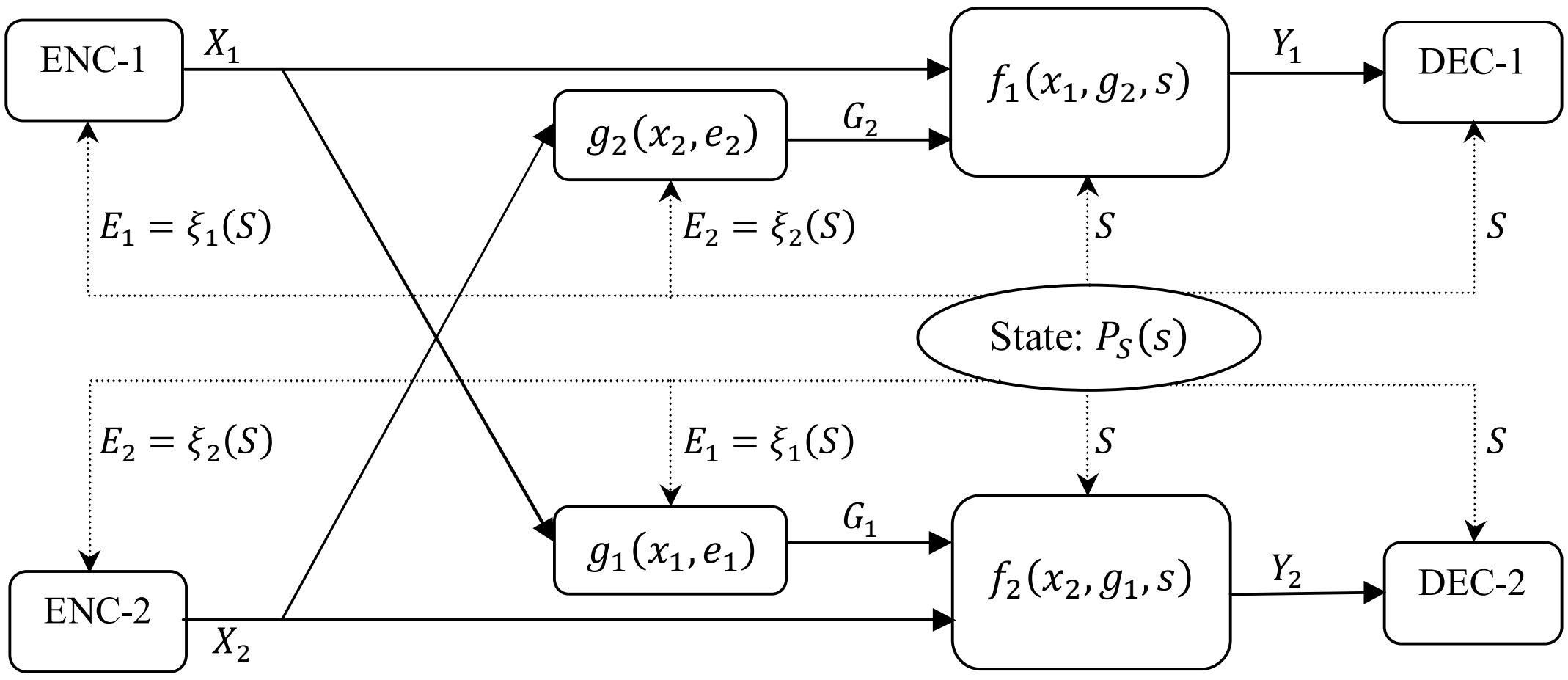

Figure 3. The state-dependent injective deterministic IC whith partial CSIT.

where $G_1 = g_1(X_1, E_1)$ and $G_2 = g_2(X_2, E_2)$ are functions of $(X_1, E_1)$ and $(X_2, E_2)$, respectively. It is assumed that the functions $f_1(.)$ and $f_2(.)$ are injective in $g_1$ and $g_2$, respectively, i.e., for every pair $(x_1, s)$, $f_1(x_1, g_2, s)$ is a one-to-one function of $g_2$ and for every pair $(x_2, s)$, $f_2(x_2, g_1, s)$ is a one-to-one function of $g_1$. Note that these conditions imply the following:

$$H(Y_1|X_1, S) = H(G_2|S), \qquad H(Y_2|X_2, S) = H(G_1|S) \tag{58}$$

Also, we assume that both receivers have access to the channel state perfectly. One can show that the capacity region of state-dependent injective deterministic IC in Fig. 3 is derived by substituting $W_1 = G_1$ and $W_2 = G_2$ in (3) (note that this substitution is adjustable with the distribution of the HK region in (3) because according to the channel definition, $G_1$ and $G_2$ depend on $E_1$ and $E_2$, respectively). We do not deal with the details for this issue as it is out of the scope of the paper.

As a natural generalization of the deterministic model, one can consider a *state-dependent semi-deterministic injective IC*. Clearly, let the functions $g_1$ and $g_2$ in Fig. 3 be stochastic in the sense that $G_1$ and $G_2$ are related to $(X_1, E_1)$ and $(X_2, E_2)$ by certain transition probability functions $P(.|x_1, e_1)$ and $P(.|x_2, e_2)$, respectively. It is clear that the state-dependent semi-deterministic injective IC includes the deterministic channel as a special case. However, in general this is not true for the Gaussian fading channel (recall that the injective semi-deterministic channel introduced in [34] always includes the Gaussian channel with fixed gains as a special case). The fact is that only when $S_{21}$ and $S_{12}$ are available at the first and the second transmitters, respectively, the Gaussian fading channel (1) becomes a special case of the state-dependent injective semi-deterministic channel with

$$\begin{cases} G_1 = S_{21}X_1 + Z_2 \\ G_2 = S_{12}X_2 + Z_1 \end{cases} \tag{59}$$

Note that the above substitution is not admissible if either $S_{21}$ is not given by $E_1$ or $S_{12}$ is not given by $E_2$ because $G_1$ and $G_2$ should stochastically depend on $(X_1, E_1)$ and $(X_2, E_2)$, respectively.

Inspired by the deterministic channel and the approach of [34], one can develop inner and outer bounds for the state-dependent injective semi-deterministic IC which differ by no more than a gap specified in terms of certain mutual information functions. This result is also applicable for the fading channel (that yields a one bit gap) where $S_{21}$ and $S_{12}$ are respectively available at the first and the second transmitters. Note that this approach yields only a special case of Theorem 2 because to derive the "one bit gap" result in Theorem 2 using the outer bound (45), it is required that the first and the second transmitters have access only to the magnitudes of $S_{21}$ and $S_{12}$, i.e., $|S_{21}|$ and $|S_{12}|$, respectively, and the availability of the phases is immaterial. The approach we followed in this paper indeed includes several benefits more. We explicitly derived an efficient outer bound for the capacity region of the Gaussian fading IC with how much state information is available at the transmitters; for example, it is also valid for the channel with no CSIT that is of particular importance from practical viewpoints. Also, our outer bound is strictly tighter than the one that is derived based on the analysis of the injective semi-deterministic channel. In fact, unlike to ours, the outer bound that results from the latter approach is not optimal for the strong interference regime (see [35, Ch. 6, Remark 6.9]).